%% file: FullSLRofMDS.tex
\documentclass[conference]{IEEEtran}
\usepackage{color}
\definecolor{orange}{RGB}{255,127,0}
\usepackage{hyperref}

\usepackage{xcolor,colortbl}

\usepackage{amssymb}
\usepackage{graphicx}

\usepackage{xcolor,colortbl}
\usepackage{booktabs} % Allows the use of \toprule, \midrule and \bottomrule in tables for horizontal lines
\usepackage{multirow}
\usepackage{tabularx} 

\usepackage{txfonts}
\usepackage{textcomp}
\usepackage{xspace}

\usepackage{lscape}
\usepackage{rotating,afterpage}

\usepackage{listings}
\usepackage{flushend}

\usepackage{subfig}

\usepackage{xspace}

\newcommand{\slr}{\textsc{SLR}\xspace} % Systematic Literature Review
\newcommand{\slrs}{\textsc{SLR}s\xspace} % Systematic Literature Reviews
\newcommand{\mds}{\textsc{MDS}\xspace} % Model-Driven Security
\newcommand{\mda}{\textsc{MDS}\xspace} % Model-Driven Architecture
\newcommand{\mde}{\textsc{MDE}\xspace} % Model-Driven Engineering
\newcommand{\mdd}{\textsc{MDD}\xspace} % Model-Driven Development
\newcommand{\mbe}{\textsc{MBE}\xspace} % Model-Driven Development
\newcommand{\soc}{\textsc{SOC}\xspace} % Separation of Concerns
\newcommand{\mof}{\textsc{MOF}\xspace} % Meta-Object Facility
\newcommand{\uml}{\textsc{UML}\xspace} % Unified Modelling Language
\newcommand{\dsl}{\textsc{DSL}\xspace} % Domain Specific Language
\newcommand{\dsm}{\textsc{DSM}\xspace} % Domain Specific Modelling
\newcommand{\aom}{\textsc{AOM}\xspace} % Aspect-Oriented Modelling
\newcommand{\aosd}{\textsc{AOSD}\xspace} % Aspect-Oriented Software Development
\newcommand{\md}{\textsc{MD}\xspace} % Model-Driven

\newcommand{\pim}{\textsc{PIM}\xspace} % Platform-Independent Model
\newcommand{\psm}{\textsc{PSM}\xspace} % Platform-Specific Model
\newcommand{\mmt}{\textsc{MMT}\xspace} % Model-To-Model Transformation
\newcommand{\mmts}{\textsc{MMTs}\xspace} % Model-To-Model Transformations (plural)
\newcommand{\mtt}{\textsc{MTT}\xspace} % Model-To-Text Transformation
\newcommand{\mtts}{\textsc{MTT}s\xspace} % Model-To-Text Transformation (plural)
\newcommand{\soa}{\textsc{SOA}\xspace} % Service-Oriented Architectures
\newcommand{\umlsec}{\textsc{UMLsec}\xspace} % UMLsec
\usepackage{graphicx}% http://ctan.org/pkg/graphicx
\usepackage{booktabs}% http://ctan.org/pkg/booktabs
\usepackage{xparse}% http://ctan.org/pkg/xparse
\NewDocumentCommand{\rot}{O{45} O{1em} m}{\makebox[#2][l]{\rotatebox{#1}{#3}}}%
\usepackage{textcomp}

\usepackage{pifont}% 
\usepackage{ifthen}
\newboolean{showcomments}
\setboolean{showcomments}{true}
\ifthenelse{\boolean{showcomments}}
 { \newcommand{\mynote}[2]{
      \fbox{\bfseries\sffamily\scriptsize#1}
        {\small$\blacktriangleright$\textsf{\textcolor{red}{{\em #2}\bf }}$\blacktriangleleft$}}}
        { \newcommand{\mynote}[2]{}}

\usepackage[%
citestyle=numeric,%
style=numeric,%
doi=false,isbn=false,url=false,% no long URLs except for online resources
firstinits=true,% abbreviate first names
hyperref,% clickable citations
maxcitenames=1,% truncate author lists in citations
maxbibnames=2,% truncate author lists in bibliography
labelnumber,%  pass number of the bibliography entry to bibliography driver for numeric citation schemes
defernumbers,% assign numeric labels the first time an entry is printed in any bibliography
sortcites=true, % sort citations in a citation if multiple entry keys are passed to the citation
backend=biber
]{biblatex}
\AtEveryBibitem{% Clean up the reference list rather than editing the entries
  \clearname{editor}%
  \clearlist{language}%
  \clearfield{month}%
  \clearlist{location}%
}
\AtEveryCitekey{% Clean up citations rather than editing the entries
  \clearname{editor}%
}

\newcommand{\footnotemarkforlastfootnote}[0]{\addtocounter{footnote}{-1}\footnotemark\xspace}

\begin{document}

%\title{An Extensive Systematic Literature Review with Snowballing on Model-Driven Security}

%\title{An Extensive Systematic Literature Review on how Model-Driven Security has been supporting the Development of Secure Systems\titlenote{This work is supported by the Fonds National de la Recherche (FNR), Luxembourg, under the MITER project C10/IS/783852.}}

%\title{An Extensive Systematic Literature Review of \\Model-Driven Security}
\title{An Extensive Systematic Review on Model-Driven Development of Secure Systems}

%\title{A Full Systematic Literature Review on how Model-Driven Security has supported Secure Systems Development\titlenote{This work is supported by the Fonds National de la Recherche (FNR), Luxembourg, under the MITER project C10/IS/783852.}}

%\author{\IEEEauthorblockN{Michael Shell}
%\IEEEauthorblockA{School of Electrical and\\Computer Engineering\\
%Georgia Institute of Technology\\
%Atlanta, Georgia 30332--0250\\
%Email: http://www.michaelshell.org/contact.html}
%\and
%\IEEEauthorblockN{Homer Simpson}
%\IEEEauthorblockA{Twentieth Century Fox\\
%\and
%\IEEEauthorblockN{James Kirk\\ and Montgomery Scott}
%\IEEEauthorblockA{Starfleet Academy\\
%San Francisco, California 96678-2391\\


\author{\IEEEauthorblockN{Phu H. Nguyen\IEEEauthorrefmark{1},
Max Kramer\IEEEauthorrefmark{2},
Jacques Klein\IEEEauthorrefmark{1} and
Yves Le Traon\IEEEauthorrefmark{1}}
\IEEEauthorblockA{\IEEEauthorrefmark{1}Interdisciplinary Center for Security, Reliability and Trust (SnT)\\
University of Luxembourg,
4 rue Alphonse Weicker, L-2721 Luxembourg\\ Email: (phuhong.nguyen, jacques.klein, yves.letraon)@uni.lu}
%Telephone: (+352) 466 644 5883
\IEEEauthorblockA{\IEEEauthorrefmark{2}Karlsruhe Institute of Technology\\
Am Fasanengarten 5, D-76131 Karlsruhe, Germany\\
Email: max.e.kramer@kit.edu}
%Telephone: (800) 555--1212, Fax: (888) 555--1212}
%\IEEEauthorblockA{\IEEEauthorrefmark{4}Tyrell Inc., 123 Replicant Street, Los Angeles, California 90210--4321}
}


\maketitle

\begin{abstract}
\textit{Context}: Model-Driven Security (\mds) is as a specialised Model-Driven Engineering research area for supporting the development of secure systems. 
Over a decade of research on \mds has resulted in a large number of publications. 
\\
\textit{Objective}: To provide a detailed analysis of the state of the art in \mds, a systematic literature review (\slr) is essential. 
\\
\textit{Method}: We conducted an extensive \slr on \mds. 
Derived from our research questions, we designed a rigorous, extensive search and selection process to identify a set of primary \mds studies that is as complete as possible. % ``most complete .. as possible'' is not possible in English
Our three-pronged search process consists of automatic searching, manual searching, and snowballing. 
After discovering and considering more than thousand relevant papers, we identified, strictly selected, and reviewed $108$ \mds publications. 
\\
\textit{Results}: 
% Max:
The results of our \slr show the overall status of the key artefacts of \mds, and the identified primary \mds studies. 
E.g. regarding security modelling artefact, we found that developing domain-specific languages plays a key role in many \mds approaches. 
The current limitations in each \mds artefact are pointed out and corresponding potential research directions are suggested.
%The current limitations of \mds approaches are pointed out and some corresponding potential research directions are suggested. 
% ``Among .. we .. discuss .. into'' is not correct and style guidelines suggest to avoid slashes (/) in normal text
Moreover, we categorise the identified primary \mds studies into $5$ principal \mds studies, and other emerging or less common \mds studies. 
Finally, some trend analyses of \mds research are given. 
\\
\textit{Conclusion}: %As a major outcome of the review, the
% As a major outcome of the review, the results strongly suggest the needs
% a) In my opinion ``As a major outcome of the review, the results .. suggest'' says nothing more than ``The results suggest'' and only makes the abstract longer
% b) need should not be used in its plural form here
% c) In my opinion it is better to say that ``our results show the need for'' than saying that they only ``suggest the need for'' as this is much weaker
Our results suggest the need for addressing multiple security concerns more systematically and simultaneously, for tool chains supporting the \mds development cycle, and for more empirical studies on the application of \mds methodologies. 
% a) Conditional should only be used if there is a condition. Here we do not have a condition. To the best of our knowledge it _is_ like this. This is stronger than it _could be_ like this.
% b) You can only use ``the'' (here before snowballing strategy) for terms or things that you already introduced. Therefore, it has to be ``a snowballing strategy''. ``The'' could be used if you want to say that there is only one snowballing stategy in the world and you want to refer to it. But, this is not the way we use it later in the paper.
To the best of our knowledge, this \slr is the first in the field of Software Engineering that combines a snowballing strategy with database searching. 
This combination has delivered an extensive literature study on \mds.
\end{abstract}



%\keywords{Systematic review, Model-driven security, MDS, Model-driven engineering, MDE, Security engineering, Security, Review, Survey, Model transformations.}
%%systematic review; survey; model-driven security; model-driven; security; model; model transformations; 

% \fi
\IEEEpeerreviewmaketitle

%\graphicspath{{Figs/PNG/}{Figs/PDF/}{Figs/}}

%%%%%%%%%%%% SECTION %%%%%%%%%%%%%%%%%%%%%%
%\input{introduction.tex}
\section{Introduction}
%Introduction
%The specific research problem or question that your thesis work will address
\label{sect.intro}
%The further the digital age progresses, the more significant role that IT security engineering would play, i.e. to protect the increasing amount of sensitive/important information being used and stored in electronic format. 
% With the progress of the digital age, IT security engineering is becoming increasingly important in building secure IT systems. 
With more and more IT systems being developed and used, approaches for systematically engineering \emph{secure} IT systems are becoming increasingly important.
\emph{Model-Driven Security} (\mds) emerged more than a decade ago as a special area of \emph{Model-Driven Engineering} (\mde) for supporting the development of secure systems. 
\mde has been considered by some researchers as a solution to handle complex and evolving software systems \cite{Bezivin2006}. 
It leverages \emph{models} and \emph{transformations} as main artefacts at every development stage. 
%\mds is the common term of scientific approaches for the model-driven development of secure systems. 
\mds specialises \mde by taking security requirements and functional requirements into account at every stage of the development process. 
%@Max: this sentence is not correct: \emph{Modelling} a desired system and manipulating these \emph{models}, raises the level of abstraction and can provide several benefits, especially regarding security engineering. 
By \emph{modelling} and manipulating models, the level of abstraction is higher than code-level that brings several significant benefits, especially regarding security engineering.
\emph{First}, security concerns can be considered together with business logic and other quality requirements such as performance from the very beginning, and throughout the \mds development life cycle. 
% In this way, security requirements are considered early, and implemented more properly. 
\emph{Second}, reasoning about systems at the model level, e.g. with model-based verification and validation methods, makes it possible to check security requirements and other requirements at early design stages. 
% Formal methods such as model checking and model-based analysis can be employed for verifying security properties. 
% Model-based security testing methods could be employed for validating the resulting secure systems (especially in where formal methods would not be applicable). 
These methods can perform formal verification as well as security testing based on models.
Moreover, models that abstract away from target platform details can increase cross-platform interoperability. 
%Security experts can therefore focus on security-related issues, instead of dealing with the technical problems of integrating those issues in the system infrastructure. 
\emph{Third}, \mds can be more productive, and supposedly less error-prone than traditional development methods by leveraging automated \emph{model-to-model transformations} (\mmts) and \emph{model-to-text transformations} (\mtts, code generation). 
%\mds focuses on making (security) models productive, i.e. enforceable in the final deployment. 
% All these points show that \mds could be the solution to all the challenges for developing modern secure systems mentioned in the first paragraph. 
%\mk{this is the first paragraph and the mentioned challenges are no longer part of the intro, therefore the sentence ``All these points ...'' is no longer part of the intro}

%In the very late 20$^{th}$ century, 
%The general concept of \mds (in its earliest forms) was found, mostly in academia, such as \cite{Epstein:1999:TUB:319171.319184}, \cite{Jurjens2002}, \cite{Lodderstedt2002}, \cite{Basin2003}. 
%From the beginning of the 21$^{st}$ century until now, many research results in \mds have been published even though \mds is a relatively young research discipline. 
For more than a decade since \mds first appeared, a considerable number of \mds publications has shown a great attention of the research community to this area. %, knowing that security getting more crucial. 
The \mds approaches vary greatly in many artefacts such as the security concerns addressed, the modeling techniques used, the model transformations techniques used, the targeted application domains, or the evaluation methods used. %, and so on. 
To provide a detailed state of the art in \mds, a full systematic literature review (\slr) is needed. 
%A systematic review is important because it summarizes existing techniques concerning a research interest, it identifies further research directions, and it provides a framework to position new research activities \cite{Kitchenham_2007}. 

So far, %related work %(\cite{Kasal:2011:MDM:1955602.1956038,Basin:2011:DMS:1998441.1998443,AdvMDS,Uzunov:jucs_18_20:engineering_security_into_distributed,Jensen:2011:SMD:2065363.2066253}) 
a full \slr on \mds does not exist. 
Surveys on \mds approaches (\cite{Kasal:2011:MDM:1955602.1956038,Basin:2011:DMS:1998441.1998443,AdvMDS,Uzunov:jucs_18_20:engineering_security_into_distributed}) could provide in-depth analyses of some well-known \mds approaches, but do not summarize the complete research area systematically. % to position new research activities. 
\cite{Jensen:2011:SMD:2065363.2066253} could be closer to our work, but has several limitations in terms of scope and methodology. 
E.g., it missed many important primary \mds approaches such as UMLsec \cite{Jurjens2002}, and aspect-oriented approaches. % (e.g. \cite{Ray2004575,Kim2006,Mouheb2010}). 
%it is still difficult to know which \mds approach(es) one should adopt/base on to address some specific security concerns. 
%Moreover, appropriate tool support is required for further adoption of \mds by the industry \cite{springerlink:10.1007/978-3-540-69100-6_31}. 
In contrast, our \slr is performed in both width and depth of \mds research that reveals an extensive set of primary \mds studies. 
Furthermore, our review provides a detailed overview on key artefacts of every \mds approach such as used modeling techniques, considered security concerns, employment of model transformations, verification or validation methods, and targeted application domains. 
Finally, we present trend analyses for \mds publications, and for the addressed security concerns and other key artefacts. 
% We give a detailed comparison of our \slr with related work in Section \ref{sect.relatedWork}. 

%This paper is an extended report of an update of a previous review \cite{Nguyen:APSEC2013}. 
This paper is an extended and improved version of \cite{Nguyen:APSEC2013}. 
In the previous version, we reported the results of a \slr based on $80$ \mds papers found from an automatic search and a rigorous selection process. 
In this extended version, we improved our set of primary \mds papers by conducting two more search strategies: manual search and snowballing. %\jk{Put this in introduction. Not Here.}
%It examines existing literature work in \mds in a systematic way, classifies and compares different approaches, underscores open issues, and suggests potential research directions. 
% Compared to \cite{Nguyen:APSEC2013}, this version reports on the results derived from a much more improved 
On the resulting set of $108$ finally selected \mds papers, 
% Furthermore, based on the extensive set of primary \mds papers,
we performed more detailed analyses for key artefacts, primary \mds studies, and trend analyses for a period of more than a decade.

The main contributions of this paper are: 
%1) an extensive three-pronged search strategy for identifying publications on \mds; 
1) detailed and condensed results on key \mds artefacts of all identified primary \mds publications; %such as the application domains, the addressed security concerns, the model transformations used
2) a diagnosis of limitations of current \mds approaches with suggestions for potential \mds research directions; 
3) a classification of principal and emerging/less common \mds approaches; 
and 4) trend analyses.

%@2Max: it sounds very boring for the readers if we use the phrase "we.. we..." repeatedly
%The remainder of this paper is structured as follows:	
%In Section \ref{sect.concepts}, we present background concepts and definitions used in this paper. 
%In Section \ref{sect.reviewMethod}, we describe the objective, research questions, search strategy, and selection process of our \slr. 
%In Section \ref{sect.evaluationCriteria}, we present our evaluation criteria and data extraction strategy. 
%In Section \ref{sect.results}, we report on the main results of our review. 
%%We give an in-depth discussion based on the results of our review in Section \ref{sect.discussion}. 
%In Section \ref{sect_Validity}, we discuss threats to validity and in Section \ref{sect.relatedWork}s we present related work.
%In Section \ref{sect.conclusion}, we conclude the paper, highlight open issues, and discuss future work. 
%%The open issues and suggestions are discussed in Section \ref{sect.discussion}, followed by the conclusion given in Section \ref{sect.conclusion}.

The remainder of this paper is structured as follows. 
Section \ref{sect.concepts} provides some main background concepts and definitions that are used in this paper. 
The objective of this \slr, its research questions, search strategy, and selection process are described in Section \ref{sect.reviewMethod}. 
In Section \ref{sect.evaluationCriteria}, we present our evaluation criteria and data extraction strategy. 
Section \ref{sect.results} shows the main results of our review. 
Threats to validity are discussed in Section \ref{sect_Validity}. 
In Section \ref{sect.relatedWork}s, we position this work regarding related work. 
Section \ref{sect.conclusion} concludes the paper by summarising the results, highlighting open issues, and giving some thoughts on future work. %and also suggesting some potential research directions?

%%%%%%%%%%%% SECTION %%%%%%%%%%%%%%%%%%%%%%
%\input{concepts.tex}
\section{Background Concepts and Definitions}
\label{sect.concepts}

%\cite{schumacher2013security} introduces several helpful concepts such as security "properties" (CIA), "services" (authentication, authorisation, accounting, auditing, non-repudiation), "approaches" (planning, prevention, detection, response), and "mechanisms" (access control, etc.).

\subsection{Systematic Literature Review and Snowballing}
\slr is a means for thoroughly answering a particular research question, or examining a particular research topic area, or phenomenon of interest, by systematically identifying, evaluating, and interpreting all available relevant research~\cite{Kitchenham_2007}. 
Well-known guidelines for conducting \slrs in software engineering were provided by \textcite{Kitchenham_2007} and \textcite{biolchini2005systematic}. 
All individual studies that are identified as relevant research contributing to a \slr are called \emph{primary} studies \cite{Kitchenham_2007}. 
% In this paper, all \textit{primary} studies are called \textit{primary} \mds studies which are published in \textit{primary} \mds papers. 
%We call those \emph{primary} \mds studies that are more common than others in terms of publications \emph{principal} \mds approaches and all other \mds studies \emph{less common} \mds approaches. 
In this paper, based on the numbers of publications and citations of \emph{primary} \mds studies, we further classify them into \emph{principal} \mds studies, and \emph{less common} or \emph{emerging} \mds studies. 

In a \slr, it is crucial to transparently and correctly identify as many relevant research papers in the focus of the review as possible. 
The search strategy is key to the identification of primary studies and ultimately to the actual outcome of the review~\cite{Wohlin2013}. 
The guidelines by \textcite{Kitchenham_2007} for \slrs in software engineering suggest to start with a database search that is based on a search string and also called \emph{automatic search} in this paper.
They also recommend complementary searches, e.g. a \emph{manual search} on journals and conferences proceedings, references lists, and publications lists of researchers in the field. 

Both automatic search and manual search have limitations~\cite{Wohlin2013}:
The former depends on the selection of databases, on database interfaces and their limitations, on the construction of search strings, and on the identification of synonyms.
The latter depends on the selection of research outlets, e.g. journals or conferences, and cannot be exhaustive. 
Therefore \textcite{Wohlin2013} proposed the snowballing search strategy as a first step to systematic literature studies. 
The key actions of the snowballing search strategy are: 1) identify a starting set of primary papers; 2) identify further primary papers using the reference lists of each primary paper (backward snowballing); 3) identify further primary papers that cite the primary papers (forward snowballing); 4) repeat steps 2 and 3 until no new primary papers are found. 
% Details of the snowballing search strategy can be found in \cite{Wohlin2013}. 
% We find that the snowballing search strategy could be a very good complementary search strategy for 
We are convinced, that the snowballing search strategy complements the automatic and manual search strategies of \textcite{Kitchenham_2007}. 
% Snowballing could help to improve the set of primary papers found in automatic search and manual search. 
In our \slr we defined and performed a snowballing search strategy that builds on the set of primary papers found in automatic and manual searches.
Details of our search strategy are presented in Section \ref{sect.reviewMethod}.

%\subsection{Modeling, Model Transformations, etc.}
%
%TODO? introduce how modeling, models, model transformations can be defined w.r.t. \mds. 
%

\subsection{A Definition of \mds}

\begin{figure}
\center
\includegraphics[width=\columnwidth]{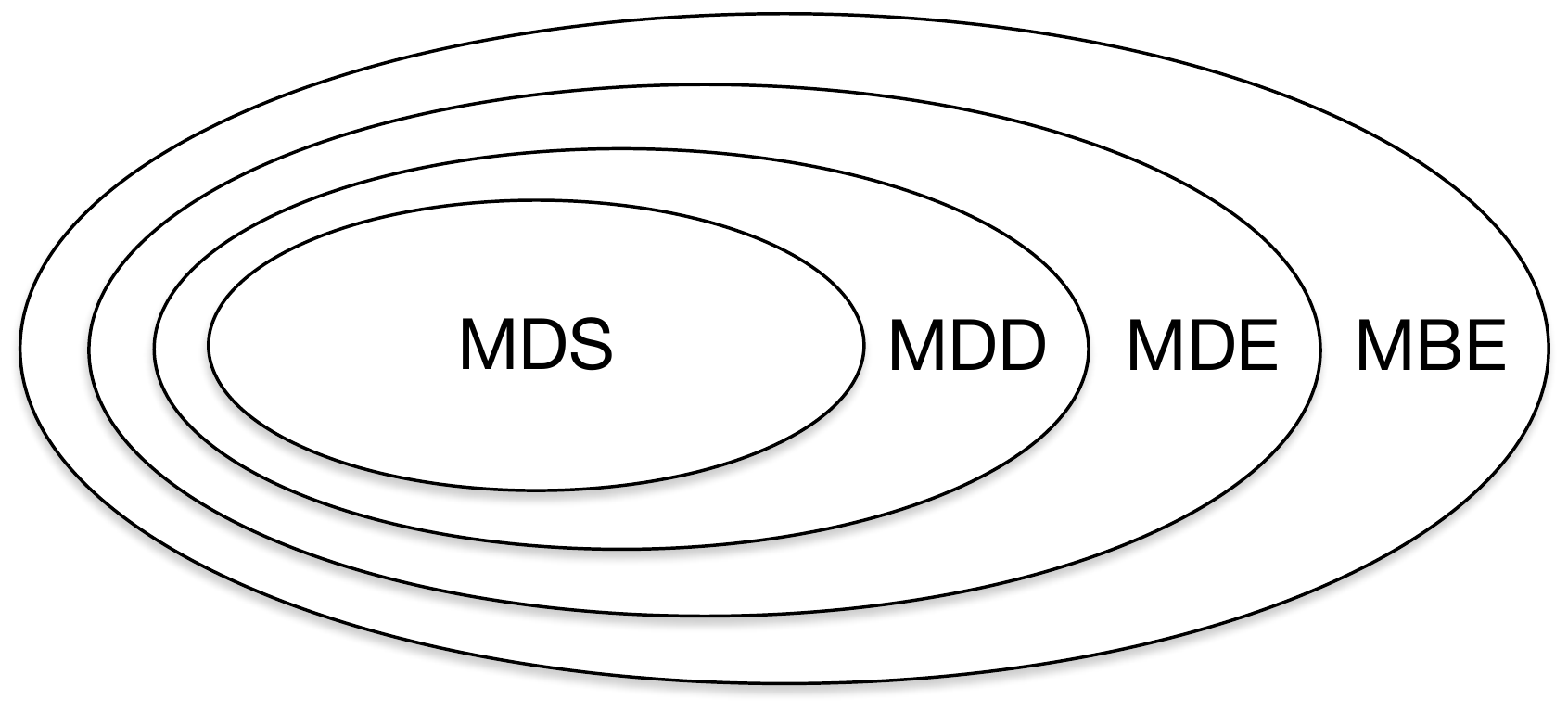}
\caption{Relations among \mbe, \mde, \mdd and \mds.}
\label{fig_Vendiagram_MD}
\end{figure}

%See the article from Jodi Cabot. 
%http://modeling-languages.com/relationship-between-mdamdd-and-mde/
%Extend it with the key points of \mds. 
%=> an inner circle of \mds within \mde. 
% Not all model-based engineering or security engineering approaches can be categorized as \mds. 
Numerous security engineering techniques exist which support the development of secure systems. 
There are also many \mde techniques for the development and maintenance of software systems in general. 
Our focus, however, is only on \mde approaches that are specifically customized for supporting the development of secure systems. 
% Here, we are not trying to define a new term but rather to give a better idea how we consider a paper as an \mds paper. 
As we already mentioned, \mds can be considered a subset of \mde. 
% To be more specific, we show how we see 
We will now clarify the relations between \mde, Model-Based Engineering (\mbe), Model-Driven Development (\mdd), security engineering, and \mds, which are important for our inclusion and exclusion criteria (Section \ref{sect.selection}). 
Regarding \mbe, \mde, and \mdd, we agree with the point of view presented by \textcite[p. 9]{BrambillaCabotWimmer201209}.
Specifically, \mbe can be used for development processes in which models may not necessarily be the central artifacts for development. 
E.g., if models are only used for documentation purposes and not in automated transformations.
% As an example given, in a \mbe development process, the platform-independent models of the system are specified by designers, but then these models are directly handed out to the programmers to manually write the code. 
% In that process, models are used in a specific step but not all steps of the development process, i.e. models are not used for (partially) automatic code generation, or those platform-independent models are not transformed into platform-specific models. 
\mde can be seen as a subset of \mbe in which models have to be the key artifacts throughout the development, i.e. models ``drive" the process in every step:
% In other words, \mde is truly model-driven in every task of a complete software engineering process:
All development, evolution, and migration tasks have to be influenced by explicit models.
\mdd can be considered a subset of \mde that only denotes development activities with models as the primary artifact. 
Normally, model-to-model transformations (\mmts) or model-to-text transformations (\mtts) are used in \mdd to obtain other models or to generate code.
Thus, \mds refers to all research approaches that focus on a \mdd process for building secure systems. 
%\mk{should we add MDD in \autoref{fig_Vendiagram_MD} between mds and mde?}
% This means that a research approach is still considered as an \mds approach if it goes beyond \mdd, but has a \mdd process in it. 
Figure \ref{fig_Vendiagram_MD} depicts these subset relations. 

%MBE > MDE > \mds > MDD
%MDA is the OMG’s particular vision of MDD and thus relies on the use of OMG standards. Therefore, MDA can be regarded as a subset of MDD.

%%%%%%%%%%%% SECTION %%%%%%%%%%%%%%%%%%%%%%
%\input{method.tex}
\section{Our systematic review method}
\label{sect.reviewMethod}

\begin{figure}
\center
\includegraphics[width=0.7\columnwidth]{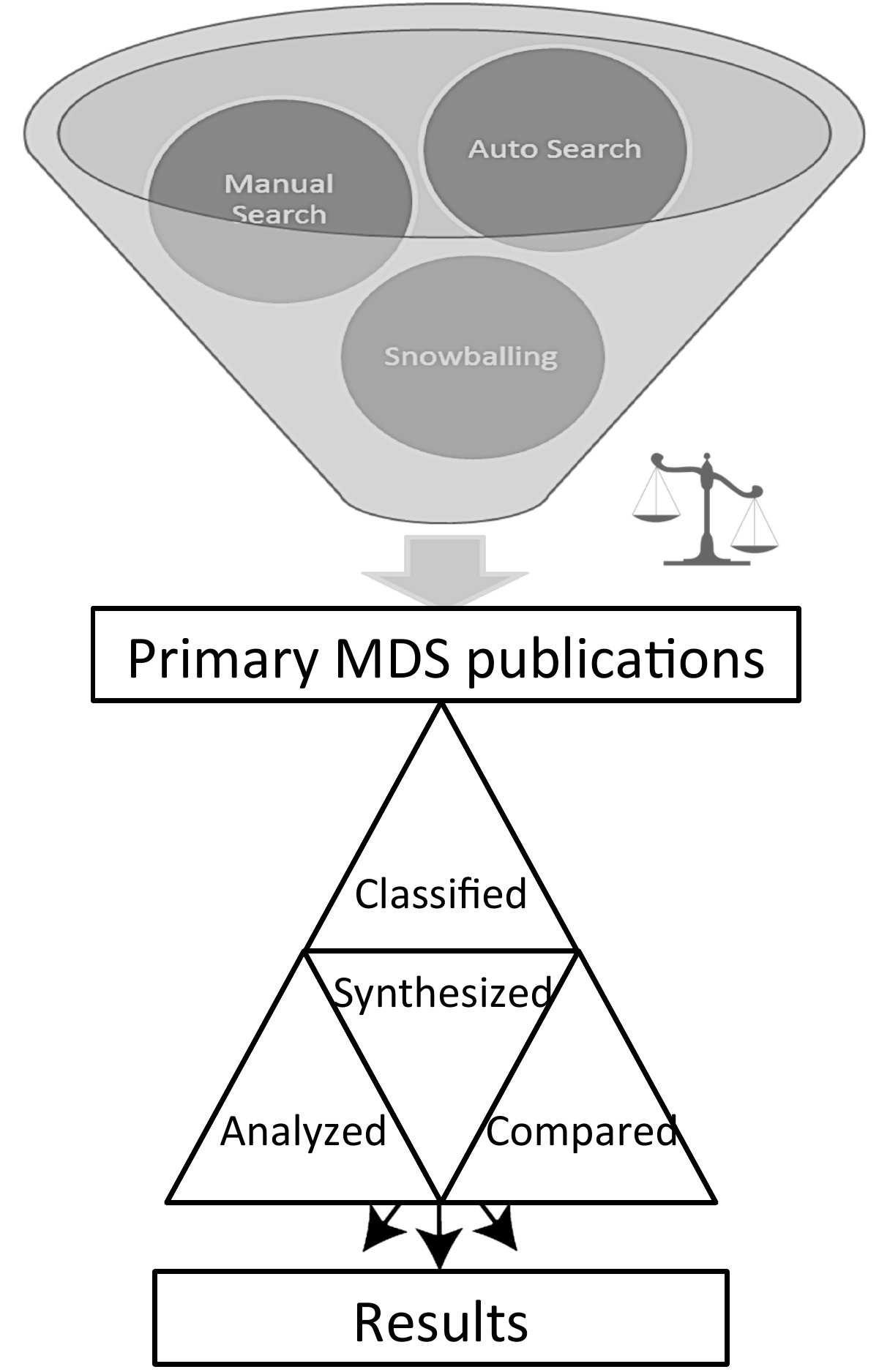}
\caption{An overview of our \slr process.} %\jk{Redraw the fig in black and white. Keep it on one Column}
\label{fig_3processes}
\end{figure}

Our \slr method is based on the guidelines of \textcite{Kitchenham_2007}, and the snowballing strategy of \textcite{Wohlin2013}. 
We presented the motivation for our review in Section \ref{sect.intro} and state our research questions in the next section. 
% Our research questions are given below (Section \ref{sect.researchQuestions}). 
Based on these research questions, we developed a review protocol, which was evaluated before conducting the review. 
Figure \ref{fig_3processes} shows an overview of our \slr process. 
We combined an automated database search (Section \ref{lbl_autosearch}), a manual search in relevant journals and conference proceedings (Section \ref{lbl_manualsearch}), and a snowballing strategy (Section \ref{lbl_snowballing}) to identify as many primary \mds papers as possible. 
For our predefined protocol we clarify the selection criteria (Section \ref{sect.selection}) to reduce a possible bias in the selection process (Section \ref{selectionProcess}). 
The quality assessment, data extraction and synthesis of the primary \mds studies are based on a fixed set of evaluation criteria (Section \ref{sect.evaluationCriteria}). 
The results obtained from classifying, synthesising, analysing, and comparing the data extracted from the primary \mds studies are presented in Section \ref{sect.results}.

\subsection{Research Questions}\label{sect.researchQuestions}
%Once the background and the motivation for conducting a \slr have been determined, the next crucial step is to define precisely the research questions that have to be answered by the end of the review. 
%In the field of \mds, it can be seen that security-oriented models are the first-class citizens and which paradigm used for modeling secure systems plays an important role in the whole development cycle. 
%In order to examine the evidence of leveraging different model-driven approaches supporting the development of secure systems, 
This \slr aims to answer the following research questions: 

\textit{\textbf{RQ1: How do existing \mds approaches support the development of secure systems? }}

This question is further divided into the following subquestions:\\
%\begin{itemize}
	RQ1.1: What kinds of \emph{security concerns} are addressed and what \emph{security mechanisms} are used by these \mds approaches?\\
	RQ1.2 \label{item_RQ1.2}: How do the \mds approaches \emph{specify} or \emph{model} security requirements together with functional requirements? Is there any tool that supports the \emph{modelling} process?\\
	RQ1.3 \label{item_RQ1.3}: How are \emph{model-to-model transformations} (\mmts) used and which \mmt engines are used? Is there any tool support for the transformation process?\\%Assuming that all the systems are developed using AOSD approaches? to transform/compose/weave security aspects into business logic (base) model?
	RQ1.4 \label{item_RQ1.4}: How are \emph{model-to-text transformations} (\mtts) used to generate code, including security infrastructure and configuration? Which tools are used for the generation process?\\
	%\item RQ1.5 \label{item_RQ1.5}: What are the different model-driven approaches used for analysing/verifying security requirements?
	%\item RQ1.5 \label{item_RQ1.5}: How model-based testing techniques are leveraged to validate the secure systems against theirs functional and security requirements?
	RQ1.5 \label{item_RQ1.5}: Which \emph{methods} were used to evaluate the approaches? What results have been obtained?\\
	RQ1.6 \label{item_RQ1.6}: Which \emph{application domains} are addressed by the \mds approaches? %Examples of application domains are information systems, Web applications, e-commerce systems, secure smart-card systems, embedded systems, distributed systems, etc.

\textbf{\textit{RQ2 \label{item_RQ2}: What are current limitations of existing \mds research?}}

\textbf{\textit{RQ3 \label{item_RQ3}: What are open issues to be further investigated?}}

%After having the research questions, the most suitable search strategy can then be employed to identify relevant studies and extract the data required to answer the questions \cite{Cai20081051}.
%All the research questions and its next steps for conducting the full review are defined clearly in our review protocol. 

%\subsection{Review Protocol}\label{sect.protocol}
%%The important role of a review protocol
%%One of the most important aspects for the success of a \slr is its well-designed review protocol. 
%%The review protocol serves as a concrete and formal scheme for conducting the \slr. 
%%By being well-designed and predefined, the review protocol ensures to reduce the possibility of reviewers' bias. 
%%\jk{Is it needed? }
%
%%How carefully our review protocol has been designed?
%Our review protocol has been formally defined and some key parts of it (e.g. the search string) were piloted for several times before included in the final protocol. 
%The protocol was initiated by one author and reviewed by the other authors. 
%We strictly followed the protocol while conducting the review. 
%Most content of our review protocol have been partially presented in Sections \ref{sect.intro}, \ref{sect.reviewMethod}, and \ref{sect.evaluationCriteria}. 
%%Readers are recommended to see our technical report \cite{} for the full (extended) protocol if they are interested. 
%Figure \ref{fig_3processes} shows an overview of our \slr process. 
%The set of primary \mds publications is obtained by conducting a rigorous selection process consisting of 3 search strategies and 8 selection criteria. 
%The results of our review are deducted from classifying, synthesizing, comparing and analyzing data extracted from the set of primary \mds publications. 
%Details of our selection process are presented in the following sections. 

\subsection{Search Strategy}\label{sect.search}
%In this section, we present the search strategy that we applied to search for relevant \mds papers. 
We developed a hybrid strategy to exhaustively search for \mds papers. 
The goal was not to miss any relevant \mds paper and therefore to find as many primary \mds papers as possible. 
Our hybrid strategy consists of three parts: automatic search (Section \ref{lbl_autosearch}), manual search (Section \ref{lbl_manualsearch}), and snowballing (Section \ref{lbl_snowballing}). 
In each step, we applied inclusion and exclusion criteria (Section \ref{sect.selection}) to select primary \mds studies.

\subsubsection{Identification of a Search String \label{sect.query}}
Based on the research questions (Sect. \ref{sect.researchQuestions}), we created search terms to form search strings, e.g. \emph{model-driven}, \emph{model-based}, \emph{security}. 
We divided our search terms into three categories: \mde (model-driven, model-based, model*, MDA, UML), modeling (specify*, design*), transformations (transform*, code generation) and security. 
%population terms, intervention terms, and outcome terms. First, the population terms are the keywords that represent the domain of model-driven security, such as security modeling, analyzing security policy, model transformations and security requirements engineering. Second, the intervention terms are the keywords that represent the techniques applied in the population to achieve an objective. In our case, the techniques for developing secure systems based on models could be very different; therefore, we decided to use general terms like transformations, code generation, and model analysis. Third, the outcome terms represent different types of security models, which could be generated. 

To form the search string, we used a conjunction that combines disjunctions of the keywords of each term group.
We had to refine our search string several times to make sure that as many potential relevant papers as possible are reached and had to adapt it according to the required format of the search engines. %In fact, the refinement process of the search string is only completed once there is no paper selected from the reference list of a selected paper (by manual search) missing from the results returned by automatic search. %???
% To be more specific, the search string is formulated as follows (but needs to be adapted for each search engine):\\
% \textit{( ``model-driven'' OR ``model based'' OR MDA OR \mde OR model* OR UML) AND ( specify* OR design* ) AND ( transform* OR ``code generation'' ) AND security} %The following groups of terms were used to form the query string:
%
%\textbf{Population terms}: security modeling; security requirements analysis; security requirements engineering; model-based analysis; model-driven analysis; analyzing security policy; secure model transformations.
%
%\textbf{Intervention terms}: modeling; model-checking; automated transformation; automatic transformation; transformation; transform; transforming.
%
%\textbf{Outcome terms}: secure model(s); UML model(s); class diagram(s); sequence diagram(s); interaction diagram(s); activity diagram(s); state machine(s); state chart(s); message sequence chart(s);

%\subsubsection{Electronic and manual search}
\subsubsection{Step 1: Automatic Search in Databases for Scientific Literature}
\label{lbl_autosearch}
Using the search string described earlier, we performed automatic search within five electronic databases for publications between 2000 and 2014: IEEE Xplore\footnote{\href{http://ieeexplore.ieee.org}{ieeexplore.ieee.org}, \href{http://dl.acm.org}{dl.acm.org}, \href{http://apps.webofknowledge.com}{apps.webofknowledge.com}}, ACM Digital Library\footnotemarkforlastfootnote, Web of Knowledge (ISI)\footnotemarkforlastfootnote, ScienceDirect (Elsevier)\footnote{\href{http://sciencedirect.com}{sciencedirect.com}, \href{http://link.springer.com}{link.springer.com}}, and SpringerLink (MetaPress)\footnotemarkforlastfootnote. 
%\mk{Do we have to keep 2012 here?}
% Each time, the search string might need to be modified to fit the format requirements of the electronic database before applying it. 

%In the previous version of this paper, we reported the results of our \slr based on the 80 \mds papers found from an automatic search and a rigorous selection process (see Table \ref{tab:stat}). 
%In this extended version, we improved our set of primary \mds papers by conducting two more search strategies: manual search and snowballing. \jk{Put this in introduction. Not Here.}

%To get those 80 \mds papers, we initiated our search by identifying a query string being used to perform electronic searches, based on our research questions in Section \ref{sect.researchQuestions}. 
%Then, we searched five electronic databases using this query string. 
%%In addition, as a complement to the electronic search, we performed manual search in specific journals and conference proceedings. 
%Next, we scanned all the sources resulting from this search to select the papers to be included in the review. 

%\begin{itemize}
%	\item IEEE Xplore \footnote{http://ieeexplore.ieee.org/Xplore/home.jsp}
%	\item ACM Digital Library \footnote{http://dl.acm.org/}
%	\item Web of Knowledge (ISI) \footnote{http://apps.webofknowledge.com}
%	\item ScienceDirect (Elsevier) \footnote{http://www.sciencedirect.com/}
%	\item SpringerLink (MetaPress) \footnote{http://link.springer.com/}
%%	\item EBSCOhost \footnote{http://www.ebscohost.com/}
%%	\item Google Scholar \footnote{http://scholar.google.com/}
%\end{itemize}

\subsubsection{Step 2: Manual Search in Conferences Proceedings and Journals}
\label{lbl_manualsearch}
To ensure the correctness and completeness of our review, we also conducted two manual searches: a manual search in potentially relevant peer-reviewed journals, and another one in potentially related conference proceedings. 
We selected journals and conferences that are highly ranked either in the domain of software engineering (SE) or security and privacy (S\&{}P). 
%\mk{We should say were the rankings come from}
%\phu{see the next paragraph}
We manually searched for all published papers from 2001 to 2014 in $10$ journals and $10$ conference proceedings as shown in Table \ref{tab:Journal_List} and \ref{tab:Conf_List}.

\begin{table*}[ht] %here
\caption{Journals used in our manual search.}
\begin{center}
    \begin{tabular}{ | l | l | l | r | }
    \hline
    \rowcolor[gray]{.9} 
    \textbf{Acronym} & \textbf{Full Name} & \textbf{Field} & \textbf{Rating}\\ \hline
    
    TSE & IEEE Transactions on Software Engineering 	&	SE	& 	56 \\ \hline      
    
    JSS & Journal of Systems and Software &	SE	 &		34\\ \hline  
    
    IEEE S\&{}P   &	 IEEE Security \& Privacy & S\&{}P 	&	31 \\ \hline
    
	TISSEC & ACM Transactions on Information and System Security & S\&{}P & 29 \\ \hline
    
	TDSC 	& 	IEEE Transactions on Dependable and Secure Computing	&  S\&{}P	&	28\\ \hline %IEEE Symposium on Security and Privacy
    
    COMPSEC & Computers \& Security &	S\&{}P	&	27\\ \hline 
    
    INFSOF	& Information \& Software Technology	 & 	SE	&	27 \\ \hline 
    
    SOSYM & Software and System Modeling	 & 	SE	&	27 \\ \hline 
     
    TOSEM & ACM Transactions on Software Engineering and Methodology & SE		&		25 \\ \hline
        
	%JCS & Journal of Computer Security 	& 	S\&{}P	&	22 \\ \hline 
      	
    ESE & Empirical Software Engineering  & 	SE	& 		20 \\ \hline 
    
    \end{tabular}
    
%    1 & IEEE Transactions on SE (TSE) & MODELS \\ \hline
%    2 & ACM Transactions on SE and Methodology (TOSEM) & AOSD \\ \hline
%    3 & Software and Systems Modeling & IEEE Symposium on Security and Privacy \\ \hline       %International Conference on Availability, Reliability and Security, SACMAT
%    4 & Transactions on Aspect-Oriented Software Development & ACM Conference on Computer and Communications Security (CCS) \\ \hline  %http://dl.acm.org/event.cfm?id=RE182&tab=pubs&CFID=141962809&CFTOKEN=56526202     
%    5 & Information and Software Technology (IST) & USENIX Security Symposium \\ \hline % https://www.usenix.org/conferences/byname/108    
%    6 & Automated SE - An International Journal & European Symposium on Research in Computer Security \\ \hline      %http://homepages.laas.fr/esorics/ 
%    7 & Software Testing, Verification and Reliability & Annual Computer Security Applications Conference \\ \hline   %http://www.acsac.org/archive/    
%    8 & Software: Practice and Experience & ICSE \\ \hline %Annual IEEE/IFIP International Conference on Dependable Systems and Networks \\ \hline  %http://2013.dsn.org/     
%    9 & International Journal on SE and Knowledge Engineering & International Symposium on
%Research in Attacks, Intrusions and Defenses \\ \hline  %http://www.raid-symposium.org/     
%    10 & Software Process - Improvement and Practice  & Computer Security Foundations Symposium \\ \hline %http://www.ieee-security.org/CSFWweb/
    
	\label{tab:Journal_List}
\end{center}
\end{table*}

\begin{table*}[ht] %here
\centering\setlength{\tabcolsep}{4pt}

\caption{Conference proceedings used in our manual search.}

\begin{center}
    \begin{tabular}{ | l | l | l | r |}
    \hline
    \rowcolor[gray]{.9} 
    \textbf{Acronym} & \textbf{Full Name} & \textbf{Field} & \textbf{Rating}\\ \hline
    ICSE   &	International Conference on Software Engineering & SE		&	60 \\ \hline
    
    CCS 	& 	ACM Conference on Computer and Communications Security		&  S\&{}P	&	54\\ \hline %IEEE Symposium on Security and Privacy
   
    S\&{}P & IEEE Symposium on Security and Privacy	 &	S\&{}P	&	49\\ \hline 
        
	USENIX & USENIX Security Symposium 	& 	S\&{}P	&	39 \\ \hline 
    
	AOSD & Modularity/Aspect-Oriented Software Development	 & SE & 37 \\ \hline
      	
    NDSS	 & 	Network and Distributed System Security Symposium	& 	S\&{}P	& 		35 \\ \hline 
    
	ACSAC & Annual Computer Security Applications Conference & 	S\&{}P	& 	29 \\ \hline      
    
	SACMAT & Symposium on Access Control Models and Technologies & 	S\&{}P	 &		28\\ \hline  
    
    ESORICS & European Symposium on Research in Computer Security	 & 		S\&{}P	&	24 \\ \hline 
     
    MODELS & Model Driven Engineering Languages and Systems	 & SE		&		21 \\ \hline

    \end{tabular}
  	\label{tab:Conf_List}
\end{center}
\end{table*}

%The 10 journals are chosen basing on the relevance and the high impact index \footnote{Journal Citation Reports 2011}, while the 10 conferences are chosen basing on the relevance and the conferences rankings provided by The Computing Research and Education 
The 10 journals are chosen based on the relevance, the high impact index (Journal Citation Reports 2011), and the field ranking in the last 10 years according to the Microsoft Research website. %\footnote{http://academic.research.microsoft.com, last access in March 2014}. 
6 journals from SE and 4 journals from S \& P were selected. 
%In the top 8 journals of the Microsoft Research website, we skipped Software Engineering Notes and IEEE Software because they are informal publication/magazines/newsletters. 
%We also skipped ENVSOFT - Environmental Modeling and Software because it is less relevant to our review. 
We added the Empirical Software Engineering journal in order to find empirical validations of \mds approaches. 
The 10 conferences are also chosen on the relevance, and the conferences field ranking in the last 10 years according to the Microsoft Research website. 
%The Computing Research and Education Association of Australasia \footnote{http://core.edu.au}

%http://core.edu.au/cms/images/downloads/conference/08sort%20acronymERA2010_conference_list.pdf
%ref: http://people.engr.ncsu.edu/txie/seconferences.htm

\subsubsection{Step 3: Snowballing for a complete set of primary \mds papers}
\label{lbl_snowballing}
The automatic search and manual search processes yielded a set $95$ primary \mds papers. 
To make sure that our final set of \mds papers is complete we adopted the snowballing strategy presented by \textcite{Wohlin2013}. 
We use the big set of primary \mds papers provided by automatic and manual searches as input for our snowballing strategy as follows. 

Figure \ref{fig_Vendiagram} shows how we formed the input set of \mds papers for snowballing. 
After conducting the automated search and applying the primary study selection procedures, we obtained a first set of $80$ \mds papers (Step 1). 
Similarly, after conducting the manual search and applying the primary study selection procedures, we obtained a second set of $29$ \mds papers (Step 2). 
We merged these two sets in order to form a set of selected \mds papers that was used for partially conducting our snowballing strategy. 
\textcite{Jalali2012} provided a comparison between the \slr method and the snowballing method. 
They state that the snowballing method can be used to complement the automated search and manual search in terms of closing the final set of primary \mds papers. 
% The core part of the snowballing strategy is conducted to complement the search results so far to make sure that the most relevant papers are found. 
Because we already performed the automatic and manual searches for obtaining a set of $95$ primary \mds papers, we only adopted the following 3 out of 5 steps of the snowballing strategy: 

\begin{figure}
\center
\includegraphics[width=\columnwidth]{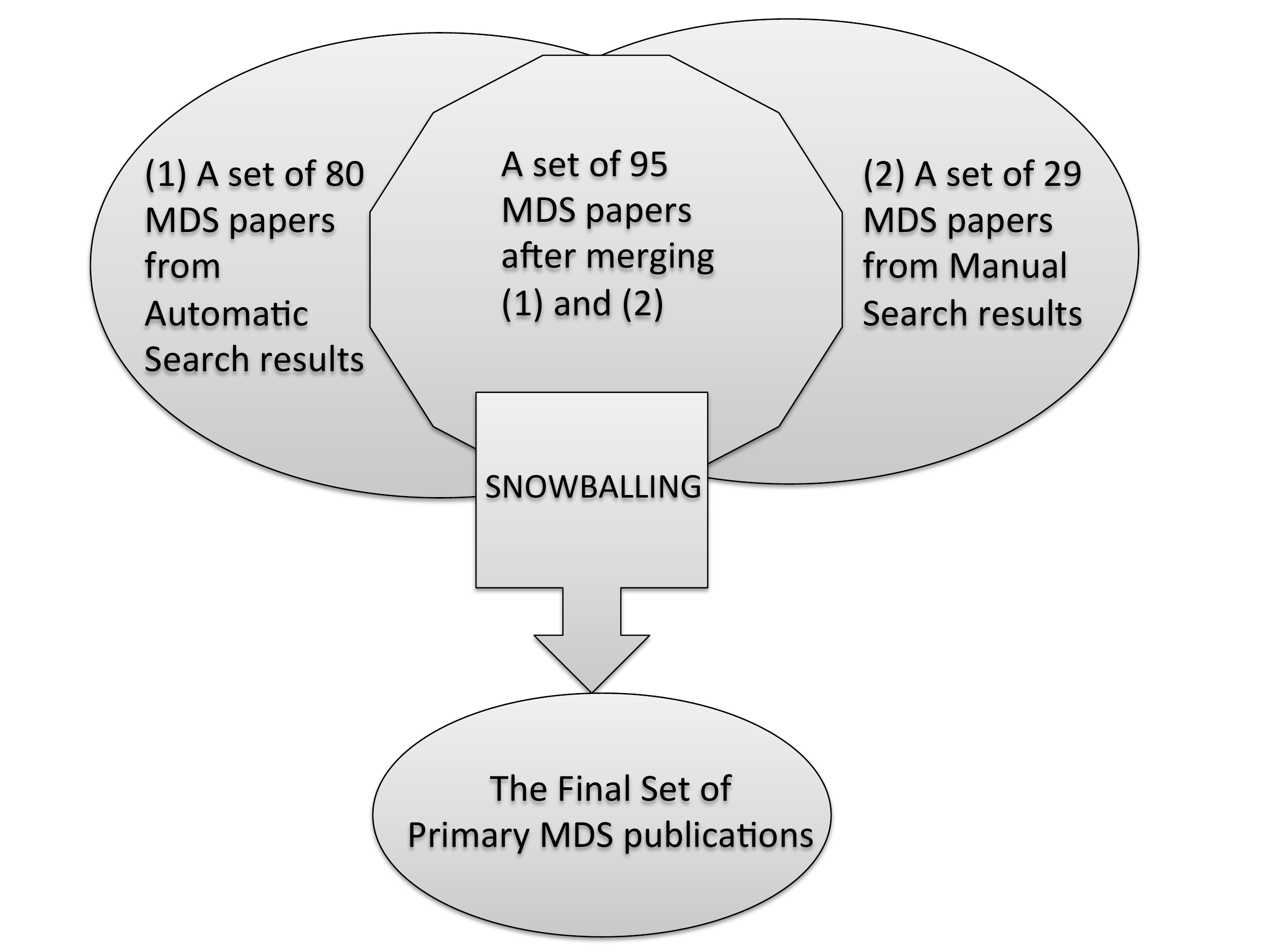}
\caption{Snowballing after Automatic Search \& Manual Search.}
%\mk{Should we add 108 between ''Final Set of`` and ''Primary MDS`` (\autoref{fig_Vendiagram})?}
%\phu{I prefer not to because we actually do the snowballing on 64 papers. Fig. 4 gives all the details}
\label{fig_Vendiagram}
\end{figure}

\begin{enumerate}
	\item \textit{Backward snowballing}: identify further potential primary \mds papers in the reference lists of the current primary \mds papers. Initially this is the set of papers found by the automated search and manual search. %,  in the reference lists to look extra thoroughly on the references of each primary \mds paper to find other . %papers by authors already included, since they obviously conduct relevant research in relation to \mds. 
% 	The backward snowballing is also used for all papers found in forthcoming steps too. 
	\item \textit{Forward snowballing}: identify further potential primary \mds papers by searching for papers that cite a current primary \mds papers. 
	We used Google Scholar%\footnote{http://scholar.google.com/}
	as recommended \cite{Wohlin2013}, because it captures more than individual databases.
	\item If no new papers are found by repeating steps 1 \& 2, then identify further primary \mds papers by searching publications lists on personal homepages or author pages of database and institutions for the primary authors of the identified primary \mds approaches.
	This step was performed to ensure that the most recent publications on the same or similar topics are included. 
%	We only check by ourselves on the homepage/google citation page but not contacting the authors of \mds papers identified to ask them for more \mds papers that they know of. 
%	Because we already conducted a rigorous combined search process including automatic search, manual search, backward snowballing, and forward snowballing, which is much less bias than involving authors' opinions. 
	%Then, we contact the authors of papers identified to ask them about: (1) papers in the area, (2) researchers conducting research in the area, and (3) additional papers from themselves. 
	If additional papers are identified then go back to Step 1.
\end{enumerate}

Once no additional papers were found in step 3, we closed the cycle of identified primary \mds papers for data extraction, synthesis, and evaluation.

\subsection{Inclusion and Exclusion Criteria}
\label{sect.selection}
We already discuss our definition of \mds to give a better idea how we consider a paper as an \mds paper in Section \ref{sect.concepts}. 
Here, we show in detail the inclusion and exclusion criteria that have been used in our primary \mds studies selection process. 

\mds approaches for developing secure system vary a great deal as different security concerns can be addressed and different model-driven techniques can be used.
Therefore, it was absolutely necessary to define thorough inclusion and exclusion criteria to select the primary studies for answering our research questions:\\ % and also conform to our research scope (Section \ref{sect.Scope}). 
% The following inclusion/exclusion criteria were used in our \slr:\\
%\begin{enumerate}
	\textit{1. Papers not written in English were excluded and already filtered out in our search process.} \\
	\textit{2. Papers with less than 5 pages in \textsc{IEEE} double-column format or less than 7 pages in \textsc{LNCS} single-column format were excluded.} \\
	\textit{3. Papers not concerned with \mde were excluded. For example, papers addressing security problems without using \mde techniques were excluded.}\\
	\textit{4. Papers proposing model-driven approaches without a focus on security concerns were excluded. E.g., model-driven approaches for performance analysis were excluded.}\\
	\textit{5. When a single approach is presented in more than one paper describing different parts of the approach, we included all these papers, but still considerd them as a single approach.}\\
	%\item When encountering more than one paper describing the same or similar approaches, which were published in different venues, we only include the most recent one of the one with the most complete description of the approach.
	\textit{6. When more than one paper described the same or similar approaches, we only included the one with the most complete description of the approach. E.g., an extended paper \cite{Nguyen:2014:TAOSD} published in a journal will be selected instead of its shorter version \cite{Nguyen:2013:MAD:2451436.2451445} published in a conference proceeding.}\\
	\textit{7. Papers with insufficient technical information regarding their approaches were excluded. E.g., papers that neither provide a detailed description of secure models, nor a precise security notion, nor transformation techniques, were considered incomplete and were excluded.}\\ % (e.g. \cite{}).
	%8. Papers not aiming at a full development cycle of secure systems are excluded. For example, papers using model-based techniques for only verifying/analyzing security mechanisms are excluded. To be more specific, we only take into account the papers using \mde approaches, i.e. the models are the central artifacts which can be used for automatic code-generation, or model transformations.
	\textit{8. Only papers with a \mdd perspectiove, i.e. \mde papers in which models are central artifacts throughout the development phase, were selected. Papers using model-based techniques only for verifying or analyzing security mechanisms without a link to the implementation code were excluded.} \\
	%\item Papers published online for more than 1 year but have been cited less than 3 times are excluded???
	%\item A paper is only included if there exists another paper discussing the same (extended) approach applied for a different stage of the development cycle. For example, 
	%\item Papers have less than 10 references are excluded???
%\end{enumerate}
\textit{9. Papers with less than 2 citations per year minus 2 as reported by Google Scholar were excluded. }

With these 9 clearly defined inclusion and exclusion criteria, we were able to perform the selection process in a more transparent and less biased way.

\subsection{Primary Studies Selection and Its Results} \label{selectionProcess}
\begin{figure}
\center
\includegraphics[width=\columnwidth]{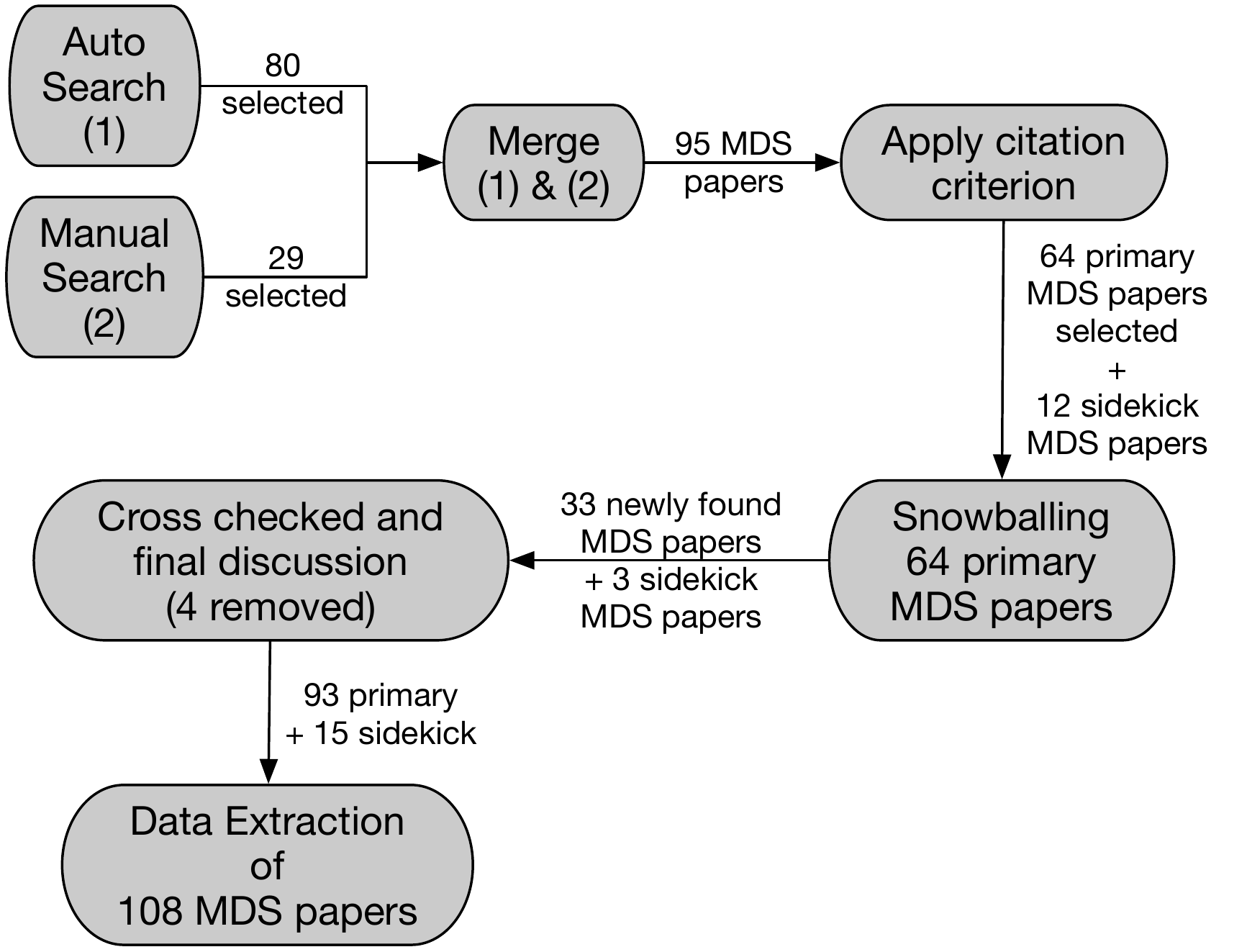}
\caption{The Selection Process with all the steps}
\label{fig_selectionProcessFinal}
\end{figure}

Here we present the selection process conducted while performing each search step in the three-pronged search process and its results. 
Figure \ref{fig_selectionProcessFinal} shows details of our whole selection process with all the numbers of \mds papers selected in each step. 

%\vspace{-8mm}
%
%\begin{figure}
%\center
%\caption{Statistic of our selection process}
%\includegraphics[width=\columnwidth]{Figs/selectionProcess}
%%\caption{Delegation Model impacting Access Control Model}%
%\label{our_overview}
%\end{figure}

%For each paper, we read the paper's title and abstract to see whether it was relevant to our research topic. 
%If the title and abstract of the paper could not help us make a decision, we further checked the paper's full text. 
%In order to augment our collection of primary studies, we scanned the reference lists of all the identified primary studies to identify additional papers. 
%Furthermore, we also went through publication lists of primary studies' authors to make sure that the most recent publications on the same or similar topics were included. 
%The statistic data of included primary studies are presented in Table \ref{tab:stat}.

\subsubsection{Selection Process in the Automatic Search Step}

\begin{table}
\centering\setlength{\tabcolsep}{4pt}

\caption{Summary of the selection process based on Automatic Search}

\begin{tabular}{>{\raggedright}lrrrrrr}
\toprule 
Source  & IEEE  & ACM  & ISI  & SD  & SL  & \textbf{Total}\tabularnewline
\midrule
Search results  & 2997  & 1506  & 3299  & 828  & 2003  & \textbf{10633}\tabularnewline
After reviewing titles/keywords  & 109  & 90  & 91  & 24  & 81  & \textbf{395}\tabularnewline
After reading abstracts  & 78  & 44  & 35  & 19  & 61  & \textbf{237}\tabularnewline
After skimming/scanning  & 31  & 21  & 17  & 15  & 20  & \textbf{104}\tabularnewline
\midrule
\midrule 
After final discussion  &  &  &  &  &  & \textbf{93}\tabularnewline
Finally selected &  &  &  &  &  & \textbf{80}\tabularnewline
\bottomrule
\end{tabular}\label{tab:stat} 
\end{table}

Table \ref{tab:stat} shows the results of our automatic search that is explained as follows. 
%\footnote{The full list of 80 finally selected papers and the links to the papers selected after reviewing titles can be found in this google project: \url{https://code.google.com/p/systematic-review-\mds/}}. 
The papers found from the repositories described in Section \ref{lbl_autosearch} were divided among reviewers. %(search engine/conference proceeding/journal) 
For each paper, we first read the paper's title, keywords, and the venue where the paper was published to see whether it is relevant to our research topic. 
If the title and keywords of a paper were insufficient for deciding whether to include or exclude it, we further checked the paper's abstract. 
If the abstract of the paper were insufficient for deciding whether to include or exclude it, we further skimmed (and scanned if necessary) the paper's full text. 
Once each reviewer had done selecting candidate papers from his repositories, all the candidate papers from different repositories were merged to remove duplicates. 
We kept track of this merging process to see which duplicates were found. 
Duplicated papers were directly included in the final set of selected papers. 
All other candidate papers, were discussed by at least two reviewers. 
Some border-line papers were checked by all reviewers. 
%First, the relevant candidates will be selected by a single reviewer. 
%Then, the candidates selected by one reviewer will be checked by another reviewer. 
We maintained a list of rejected candidate papers, with reasons for the rejection, after discussion among reviewers. 
In the end, 80 \mds papers were selected.

\subsubsection{Selection Process in the Manual Search Step}
%Similarly, our selection process in the manual search was conducted. 
29 candidate \mds papers were found in the manual search step. 
By merging with the set of 80 papers above, we obtained in total 95 \mds papers.

% \lstinputlisting[label=listing_Snowballing, caption=Our selection process while snowballing, float=*]{listings/listing_Snowballing.tex}
\begin{figure}
\center
\includegraphics[width=\columnwidth]{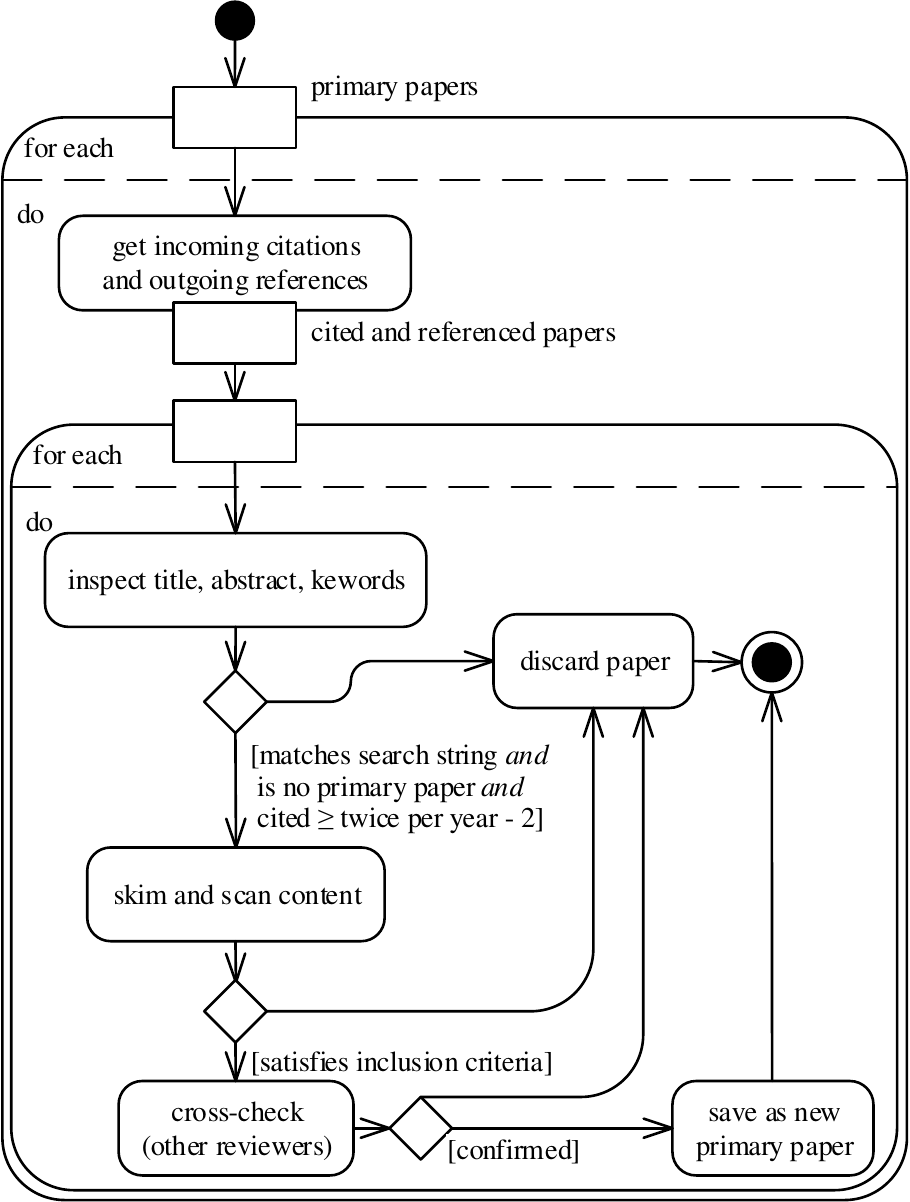}
\caption{Our selection process while snowballing}
\label{fig_snowballing_process}
\end{figure}

\subsubsection{Selection Process in the Snowballing Step}
After the first two steps, we conducted the snowballing as described in Section \ref{lbl_snowballing}. 
However, once obtaining all the numbers of citations of every paper in the set of 95 \mds papers above, we found out that some papers are much less cited than others, or even having no citation at all. 
We argue that the papers without a minimum number of citations after getting published for a specific period could be considered as not significant in terms of research impact and continuation. 
On the other hand, we also were not too strict on this aspect. 
Specifically, we decided that papers with the number of Google Scholar citations\footnote{The citations of these 95 \mds papers were dated on May 19, 2014} less than 2 citations per year minus 2 are excluded. 
Thus, the selection criterion 9 about number of Google Scholar citations was added. 
This means we leave out the papers that are not active, and do not have a minimum impact after being published for more than 2 years. 
%\mk{I suggest to remove most of this paragraph. We already presented the citation criterion above. The only new information here is the rationale (''We argue``) and the impact for papers of 2013 and 2014}
%\phu{I think it is important to explain why we added the arguable criterion 9}
Of course, this also means the recent \mds papers published in 2013 and 2014 are not excluded by this citation criterion. 

In 95 \mds papers, 31 papers were removed according to this citation criterion. 
Consequently, we used 64 primary \mds papers as the input for our snowballing process.  
In the snowballing step, we also apply the citation criterion\footnote{The citations of \mds papers found in snowballing were dated on-the-fly.} together with other criteria to select primary \mds papers. 
Details of our selection process while snowballing are %given in the pseudo code in Listing \ref{listing_Snowballing}. 
shown in \autoref{fig_snowballing_process}.
It is also important to note that every \mds candidate paper is cross-checked by three reviewers before any inclusion or exclusion decision. 
After all three steps, we have ended up with 93 primary \mds papers. 
%\mk{Are these numbers still correct?}
However, we realised that some \mds papers, which were removed because of the citation criterion, should be put back in the final set as ``sidekick" \mds papers. 
The main reason is that those \mds papers contain extra details of the approaches presented in the selected primary \mds papers. 
A ``sidekick" \mds paper is a true \mds paper that was only excluded because of the citation criterion. 
Every ``sidekick" \mds paper is part of a primary \mds approach. 
If they were removed, some important properties of the relevant primary \mds approaches could be missing in the data analysis. 
E.g., a paper presents an empirical study of a primary \mds approach. 
We would miss that empirical study of the primary \mds approach if the ``sidekick" paper was removed because of the citation criterion. 
Thus, 15 ``sidekick" \mds papers were put back in the final set. 
In the end, the final set of 108 \mds papers is used for data extraction and evaluation. 
%\mk{Can't we come up with a better name than sidekick? Sounds like Kung-Fu fighting ...} 
%\phu{Sidekick is a perfect word for it: as defined in Dictionary, sidekick is a person's assistant or close associate, especially one who has less authority than that person.}

%%%%%%%%%%%% SECTION %%%%%%%%%%%%%%%%%%%%%%
%In this section, the evaluation criteria used to evaluate selected approaches/papers is presented.
%\input{evaluationCriteria.tex}
\section{Evaluation criteria \& Data extraction strategy}
\label{sect.evaluationCriteria}%Data extraction strategy \& Evaluation criteria

%\subsection{Data extraction and quality criteria}
%a data extraction template (a spreadsheet) to col- lect the information needed to answer the research questions.
%...
%further extracted data for publication quality assessment. 
%For this, we defined the following criteria:
%Abstraction Level
%Validation Method (Industrial CS validation)
%Tool Support

%Description of data extraction forms and procedures for data extraction.
%In this section, the evaluation criteria used to evaluate selected approaches/papers is presented.
Classifications and taxonomies are important in any research domain, e.g. \cite{crnkovic2011classification}, \cite{mens2006taxonomy}. 
In this section, we describe a set of key artefacts of \mds that forms a so-called evaluation taxonomy of \mds. 
We derived our evaluation taxonomy from our research questions. 
Moreover, our evaluation taxonomy are also based on the synthesis of evaluation criteria described in \cite{doi:10.1142/S0218194002001062} and \cite{Kasal:2011:MDM:1955602.1956038}. %the evaluation taxonomy proposed in \cite{Kasal:2011:MDM:1955602.1956038}. 
Having an evaluation taxonomy makes it more systematic to assess key artefacts of \mds as well as classify and compare different \mds approaches.

Our taxonomy of \mds classifies different dimensions that one has to take into account while leveraging \mde techniques for developing secure systems. 
The elements of our taxonomy are described as follows. 
For each element, the data extraction strategy is described to show how we extracted data from the primary studies to answer our research questions. 

% model-driven techniques supporting of the development of
\textbf{Security concerns}: In this dimension, we classify primary studies according to the security concerns/mechanisms that the \mds approaches are dealing with. 
The range of security concerns is broad, e.g. authorisation, authenticity, availability, confidentiality, integrity, etc. %(Access Control, Delegation, Obligation, etc.), Role Based Access Control - RBAC, MAC
%(cryptography)
We will count the number of papers addressing each security concern. 
Thus, security topic areas that addressed by the \mds approaches are measured quantitatively.

\textbf{Modelling approaches}: Security concerns can be modelled separately or not from the business logic, and by using different modelling techniques/languages. 
%In this work, we are interested in finding how the current approaches model security concerns that can be eventually enforced into the system. 
Primary studies can be classified by the paradigms of modelling, i.e. \emph{Aspect-Oriented Modelling} (\aom) or non-\aom. 
In \aom approaches, security concerns are modelled in separate \emph{aspect models} to be eventually woven (integrated) into the \emph{primary model(s)}. 
Using \aom, security concerns can be modelled separately, modularly in design units (aspects) \cite{simmonds2005aspect}. 
Vice versa, in non-\aom approaches, security concerns are not modelled as \aom aspects. 
That means security concerns can be modelled together with business logic in every place where they are needed. 
But, we also classify as non-\aom approaches where security concerns modelled separately (\emph{separation of concerns}) from the business logic that can be integrated later into the system. 
E.g., a non-\aom approach could (separately) specify an access control policy using a \emph{Domain-Specific Language} (\dsl)\footnote{http://martinfowler.com/books/dsl.html}, and then transform and generate \textsc{XACML}\footnote{\emph{extensible Access Control Markup Language}, a XML-based declarative access control policy language} standard file for enforcing the access control policy. 
In other words, we would like to know the percentage of non-\aom approaches compared to the percentage of ``full'' \aom/\emph{Aspect-Oriented Software Development} (\textsc{AOSD}) approaches. % where security concerns are really modelled as \aom aspects. 
Separation of concerns can be considered as a key principle to cope with modern complex systems. 
Furthermore, approaches are also classified by the modelling languages, e.g. \uml diagrams, \uml profiles, or some kinds of \dsl{}s, used to model security concerns and business logic. 
The outcome models are classified as of type standard or non-standard, and structural, behavioural, functional or other types. 
The granularity levels of outcome models are also reviewed. % to see how detailed the models are. %TODO: explain clearer!

\textbf{Model-to-model transformations (\mmts) \& tools}: %\mmts can be considered as the heart and soul of \mde \cite{1231150}. 
\mmts can take part in the key steps of the development process, e.g. for composing security models into business models or transforming \emph{platform-independent models} (\pim{}s) to \emph{platform-specific models} (\psm{}s). 
We extract data related to \mmts for answering the following questions: How well-defined are the \mmts rules? 
How \mmts are implemented? 
Using which \mmt engines (e.g. \textsc{ATL}\footnote{http://www.eclipse.org/atl/}, \textsc{QVT}\footnote{http://projects.eclipse.org/projects/modeling.mmt}, \textsc{Kermeta}\footnote{www.kermeta.org}, Graph-based \mmts, etc.)? 
Is there any tool support for the transformation process? 
What is the automation level of \mmts: \emph{automatic} (if entire process of creating the target model can be done automatically), \emph{semi-automatic}, and \emph{manual}. 
Some information about the classification of \mmts should also be extracted to see if it supports well for the security mechanisms? E.g., \emph{endogenous} \mmts or \emph{exogenous} \mmts used?
%Any support for traceability w.r.t transforming security aspects?

\textbf{Model-to-text transformations (\mtts, code and security infrastructure generation) \& tools}: \mde also supports the development of secure systems by automatically generating code, including (partial) complete, configured security infrastructures. 
Data should be extracted to see the main purposes of using code generation techniques. 
Is the whole system including security infrastructure generated? 
Or just the security infrastructure (configuration) is generated? 
Can fully code and security infrastructure be generated?
Or just the (code) skeleton of the system is generated? 
Which tools are used for the code generation process? 

\textbf{Application domains}: \mds approaches are also classified on the target application domains of the secure systems. 
Some \mds approaches might target only a specific application domain. 
Some might explicitly be applicable to different application domains in general. 
Others might implicitly be applicable to different application domains. 
Some examples of application domains are information systems, web applications, databases, secure smart-card systems, embedded systems, distributed systems, etc. 
The application domains might be overlapping but could show relatively the intended application domain(s) of a specific \mds approach. 

\textbf{Evaluation methods}: To point out the limitations of each approach, we check again how the approach has been evaluated. 
How many case studies have been performed? 
What results have been obtained? 
What other evaluation methods (other than case studies) have been applied to evaluate these approaches? 
This can be answered by extracting data from the validation section of each paper. 

%Moreover, not only extracting main data to answer our research questions, we also extracted some meta-data which could reflect a big picture of \mds. 
%This set of information contains meta-data like the title, number of authors, source and year of publication, number of pages (in one-column or two-column format), number of references, number of citations. 

To make the data extraction consistent among the reviewers, we all tried to extract the relevant data from a small set of prospective primary papers. 
We then discussed to ensure a common understanding of all the extracted data items and refined the data extraction procedure. 
Excel files were used for storing the extracted data while a tool called Mendeley\footnote{\url{http://www.mendeley.com/}} was used in reviewing and controlling the selected papers. 
The final set of primary studies (selected papers) was divided among reviewers. 
Each reviewer examined again the allocated papers and enriched the Excel files to ensure detailed data according to the taxonomy has been extracted from the selected papers. 
The data extraction forms of each reviewer were read and discussed by two other reviewers. 
All ambiguities were clarified by discussion among the reviewers. 

To answer the last two research questions, we reviewed the range of security topics, the scope of \mds research work and the quality of \mds research results to determine whether there are any observable limitations and open issues. 
%Some limitations can be explicitly written in the conclusion section by authors?

%%%%%%%%%%%% SECTION %%%%%%%%%%%%%%%%%%%%%%
%\input{results.tex}
\section{Results}
\label{sect.results} %This section presents the detailed analysis and comparison of the selected related works.

\begin{figure*}[htbp]
\subfloat[How much each concern is addressed in \mds?]{\includegraphics[scale=0.55]{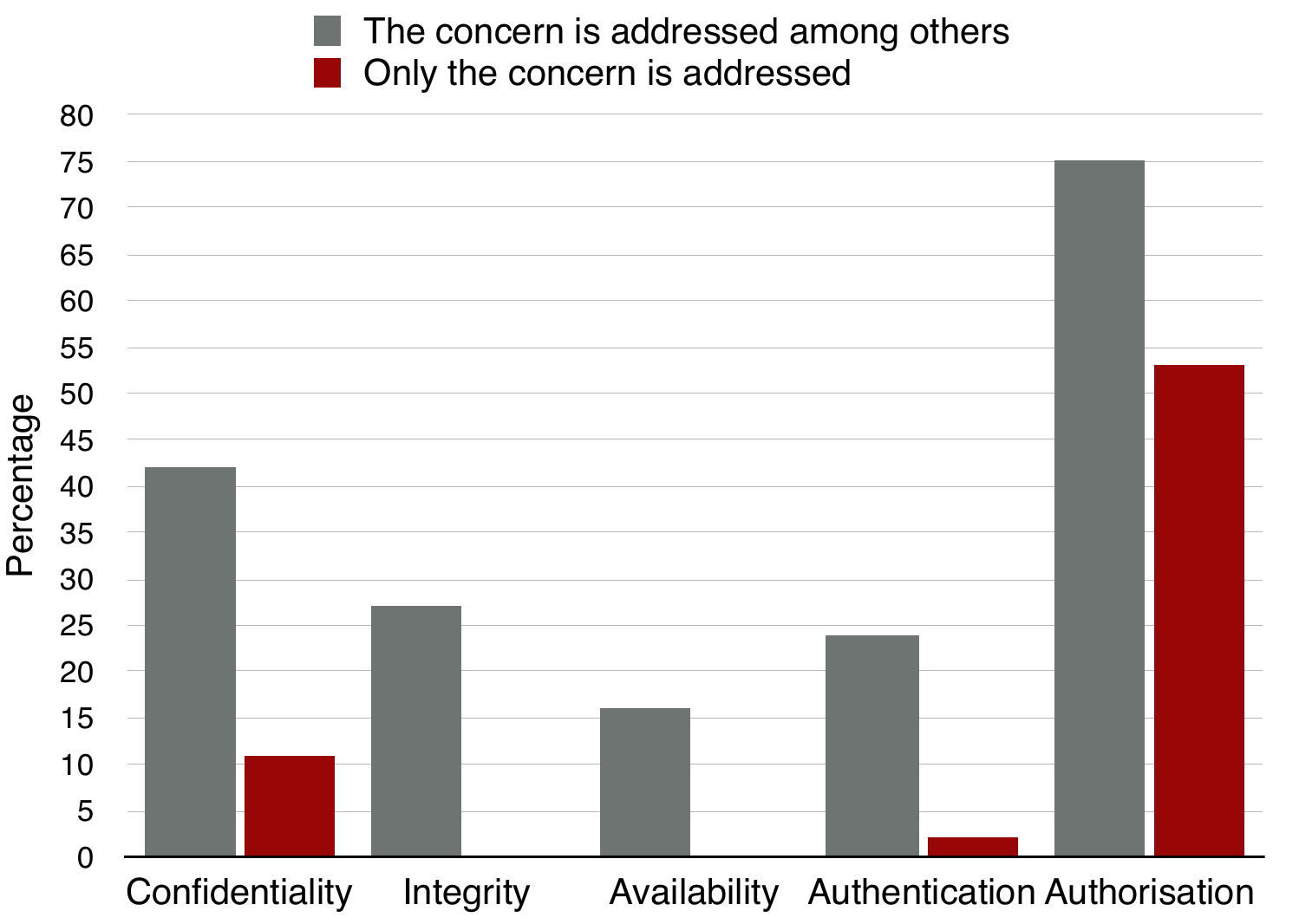}}
\subfloat[Intersection of Authentication, Authorisation, and Confidentiality]{\includegraphics[scale=0.4]{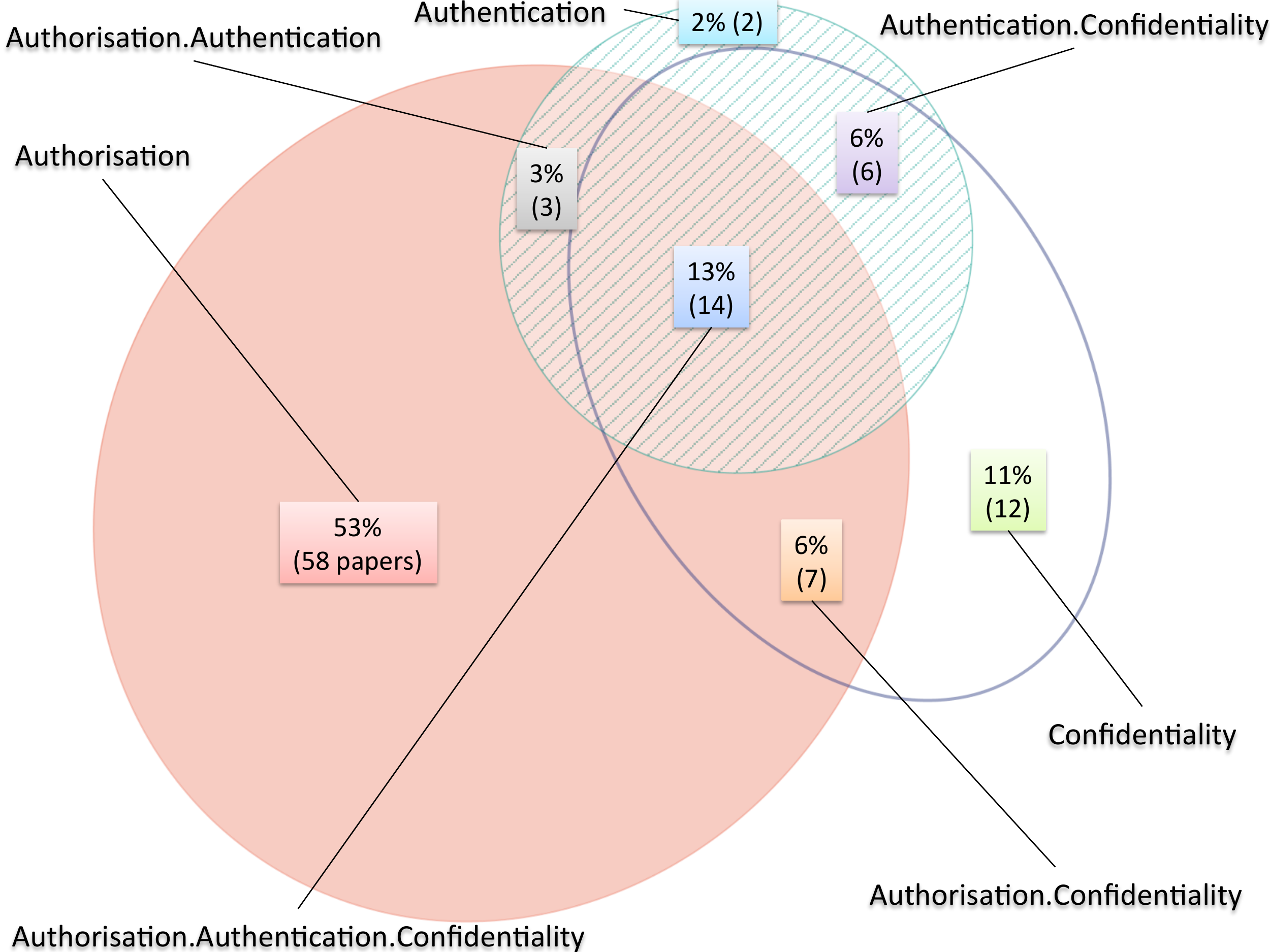}}
\caption{Statistics of Security Concerns addressed by the reviewed \mds studies}
\label{fig:secConcerns}
\end{figure*}

%\begin{figure*}
%\center
%\caption{Statistic about Security Concerns addressed by MDS}
%\includegraphics[scale=0.5]{Figs/SecurityConcerns_2015}
%\includegraphics[scale=0.5]{Figs/5star-VenDia}
%%\caption{Delegation Model impacting Access Control Model}%
%\label{fig:secConcerns}
%\end{figure*}

\begin{figure*}
\center
\includegraphics[scale=0.5]{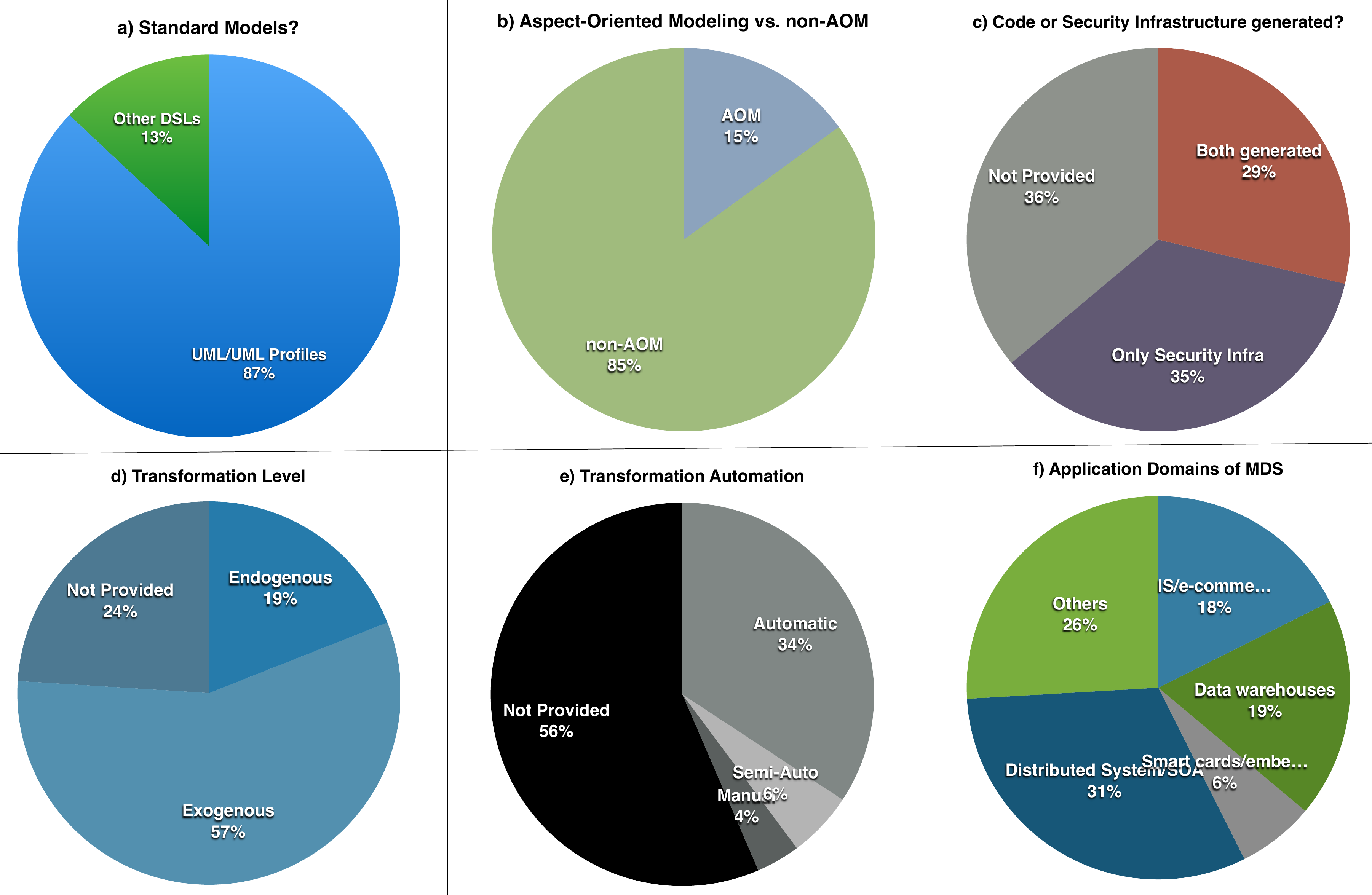}
\caption{Statistics of some key \mds artefacts}
\label{fig:piecharts}
\end{figure*}
%	\vspace{-4mm}

%By extracting and synthesising data according to the evaluation criteria presented in Section \ref{sect.evaluationCriteria}, the research questions from the beginning of this paper (in Section \ref{sect.reviewMethod}) can now be answered. 
%This section shows the results after synthesising and analysing the data extracted from the final set of selected papers. 
First, in Section \ref{sec_perCriterion} we report on some statistic results according to the evaluation criteria. 
Then, the principal \mds approaches and other emerging/less common \mds approaches are revealed and described in Sections \ref{sec_princialMDS}, \ref{sec_otherMDSApproaches} respectively. 
Finally, Section \ref{sec_trend} analyses the trends of some key factors in \mds. 

%\begin{center}
%\begin{tabular}{p{0.5\columnwidth}p{0.5\columnwidth}}
%\includegraphics[width=0.45\columnwidth]{Figs/AOMvsNonAOM} 
%& \includegraphics[width=0.45\columnwidth]{Figs/AOMvsNonAOM}
%\end{tabular}
%\end{center}

\subsection{Results per Evaluation Criterion}
\label{sec_perCriterion}

%Here the results per each evaluation criterion are presented. 
An overview of the results can be seen in Figures \ref{fig:secConcerns}, \ref{fig:piecharts} and Table \ref{tab:pubsources}. 
Fig. \ref{fig:secConcerns} shows the statistics about how each security concern has been addressed by the primary \mds approaches. 
Fig. \ref{fig:piecharts} visualises other key results for a representative set of evaluation criteria. 
Table \ref{tab:pubsources} summarises all the values for all evaluation criteria. %, e.g. number of papers/criterion.% and the publication source. 
We present the results for each evaluation criterion as follows. 

%begin{figure*}
%\begin{subfigure}{\columnwidth}
%\includegraphics[scale=0.5]{Figs/SecurityConcerns_2015}
%\caption{Each concern} \label{fig:secConcerns_bar}
%\end{subfigure}
%\hspace*{\fill} % separation between the subfigures
%\begin{subfigure}{\columnwidth}
%\includegraphics[scale=0.5]{Figs/5star-VenDia}
%\caption{Intersection} \label{fig:secConcerns_ven}
%\end{subfigure}
%\caption{Statistic about Security Concerns addressed by MDS} \label{fig:secConcerns}
%\end{figure*}

\textbf{Security concerns/mechanisms}: 
Fig. \ref{fig:secConcerns} shows the statistic of security concerns tackled by the reviewed \mds approaches. 
We can see that \emph{authorisation} is addressed the most, by 75\% of the examined \mds papers. 
Moreover, more than half of the \mds papers (53\%) deal with \emph{authorisation} only (Fig. \ref{fig:secConcerns}a,b). 
The second security concern in terms of receiving attention is \emph{confidentiality} addressed by 42\% of the examined \mds papers. 
11\% of the examined \mds papers tackle \emph{confidentiality} solely (Fig. \ref{fig:secConcerns}a,b). 
Other security concerns, like \emph{integrity}, \emph{authentication}, and \emph{availability} are, however, less tackled with 27\%, 24\%, and 16\% correspondingly. 
These results suggest that more \mds research work should focus on particular security concerns like \emph{integrity}, \emph{availability}, and \emph{authentication}. 
%There are considerable less papers tackling these security concerns than papers dealing with \emph{authorisation} and \emph{confidentiality}. 

We also would like to know how much multiple security concerns are tackled at the same time by the \mds approaches. 
Fig. \ref{fig:secConcerns}b displays the statistic about how much three key security concerns (Authentication, Authorisation, and Confidentiality) are tackled solely and simultaneously.  
Only 13\% of the examined \mds papers propose methodologies to tackle all three together. 
About 15\% of the examined \mds papers deal with two concerns simultaneously: Authentication and Authorisation (3\%), Authentication and Confidentiality (6\%), Confidentiality and Authorisation (6\%). 
Not only multiple security concerns are less tackled, but also rarely the inter-relations among multiple security concerns are formally taken into account in the reviewed \mds approaches. 
%Moreover, many \mds approaches have not dealt with multiple security simultaneously. 
Future \mds approaches should address multiple security concerns simultaneously, systematically by formally specifying inter-security concern relations. 
%Regarding modeling approaches, 
The inter-relation among security concerns have to be taken into account while developing \dsl{s} for specifying security requirements.

These first results are very interesting. 
Indeed, an open question is ``why in \mds \emph{authorisation} and \emph{confidentiality} got more attention?''. 
%In other words, why integrity, availability, and authentication are not yet extensively addressed in \mds? 
A possible answer could be that \mds is a relatively young research area with more ``model-driven" than ``security". 
\mds is the common name of the \mde approaches specifically focusing on secure systems development. 
Thus, among the authors of the published \mds papers, there are significantly more researchers with \mde background than security engineering background. 
Researchers that mainly work with \mde techniques may first address \emph{authorisation} (e.g. AC) because it is closer to application logic and functional requirements than other security concerns. 
This could be linked to the nature of security concerns. 
Some security concerns (e.g. \emph{authorisation}) are closer to the application level than others. 
\mde researcher might not be familiar with security concerns to be addressed at the network layer. 
% For people have been working mainly on \mde, when they target to solve a security problem, authorisation (e.g. access control) is often chosen because of its closeness to application logic. 
Given the background of the authors of the most renowned \mds approaches, it might be that we need more interest in \mde from the security engineering community to see more \mds approaches dealing with security concerns like \emph{integrity}, \emph{availability}, and \emph{authentication}. 
% In the opposite direction, people working mainly on security engineering would need to be more aware of \mde so that they could leverage \mde in their work.
Therefore, we suggest that more effort should be put into communicating \mde techniques as well as \mds approaches to the security engineering community. 
%In this way, more research initiatives can focus on dealing with security concerns like \emph{integrity}, \emph{availability}, and \emph{authentication}. 
% If this would be the case then other very specific security concerns like integrity, availability, and authentication could be better dealt with. 
% Anyway, there should be more effort for investigating security concerns like integrity, availability, and authentication. %, and compare them with e.g. access control. 
% There should be more effort for investigating other security concerns like integrity, availability, and authentication. 

\textbf{Modeling approaches}: 
Fig. \ref{fig:piecharts}a shows that 87\% of the examined papers used standard \uml models and defined \dsl{}s for security concerns using the profile and stereotype mechanisms of the \uml. 
13\% used other \dsl{}s (e.g. \cite{Morin2010}, \cite{Menzel2009}, or \cite{Mouelhi2008}). 
%\uml profiles were defined very often as \dsl{}s for specifying security concerns. 
Thus, we understand that standardised, common \uml models are broadly used by \mds approaches. 
On the other hand, defining \dsl{}s (either \uml profiles or other \dsl{}s) is also very popular to leverage \mde techniques for secure systems development. 
UML profiles and other kinds of \dsl{}s have been developed to better capture the specific semantics of security concerns. 
In other words, defining \dsl{}s plays a key role in \mds because that way allows expressing security concepts/elements more easily. 
%Thus, it can be seen that using \dsl{}s is more popular than using \aom approaches to leverage \mde techniques for secure systems development. 
However, using \uml profiles is not the only way for developing \dsl{}s in \mds approaches. 
\dsl{}s which are not \uml profiles are also recommended, especially \dsl{}s that can deal with multiple security concerns in the same system. 

15\% of the papers discuss approaches that are based on \aom (Fig. \ref{fig:piecharts}b) where security concerns are specified as aspects and eventually woven into primary models. 
Even though the remaining 85\% are not really aspect-oriented, most of them still follow the \emph{separation of concerns} principle and really separate security concerns from the main business logic \footnote{Note that in this paper we only classified a modelling approach as \aom if a concern is modelled as an aspect model that can be woven into a primary model. We explained this point in Section \ref{sect.evaluationCriteria}.}. 
In most of the cases, security concerns were specified separately from the business logic in \pim{}s and transformed into \psm{}s that can be refined into security infrastructures (e.g. \emph{XACML}) integrated with the systems. 

% Security concerns are often well specified by a \dsl specific for each concern. 
Security concerns are often modelled and analysed with a \dsl that is concern-specific. 
But, few \mds papers have well-defined semantics for their languages so that these languages can be used for formal analysis.
Only some papers related to the \textsc{UMLsec}, \textsc{SecureUML} approaches (see Section \ref{sec_princialMDS}) provide some formal basis for security analyses. 
This shows that further efforts are required to mature security-specific modelling languages to foster analyses. 
Most (89\%) of the \mds papers use structural models. 
Behavioural models are used in 31\% of the reviewed \mds papers. 
Other types of models like domain specific models accounted for 13\%. 
Using solely one type of models could not be enough to be able to express multiple security concerns. 
Thus, very few modelling approaches propose to deal with multiple security concerns together like \cite{Sanchez2010, Georg2009}. 
Most of them are specific to address only one security concern solely.

\textbf{Model-to-model transformations (\mmts) \& tools}: Table \ref{tab:pubsources} shows that 74\% of the papers clearly mentioned \mmts while 26\% did not use or mentioned transformations, e.g., because of a manual integration of security. %, or directly generating security infrastructures. 
More specifically, 57\% of the examined papers use exogenous transformations. 
Most of these were used to transform \pim{}s to \psm{}s (Fig. \ref{fig:piecharts}d). 
Security concerns were modelled using \dsl{}s for each concern to obtain \pim{}s that were transformed into \psm{}s, which can be refined into code. 
19\% define endogenous \mmts that are used to weave/compose security models into base models defined using the same \dsl{}s. 

34\%  of the examined \mds papers implement automatic \mmts, 6\% describe semi-automatic (interactive) \mmts, and only 4\% are manual (Fig. \ref{fig:piecharts}e). 
But 56\% do not specifically provide any implementation information about \mmts, e.g. some simply provide mapping rules for transforming models.  
Having automated \mmts is one of the key success factors of \mde \cite{hutchison2011a} and \mmts play a crucial role in \mds as well. 
Especially some important semantics of security mechanisms might be embedded in the \mmts. 
Providing \mmts implementation details in \mds is important to evaluate the efficiency of each approach. 
It can be also helpful for other researchers to learn from previous experiences in choosing or developing a suitable transformation engine for their work. 
19\% of the selected \mds papers describe their \mmts implementation using standard transformation languages like \textsc{ATL} and \textsc{QVT}. 
81\% of the papers only describe the transformation rules without implementation details, or use other transformation languages like graph-based transformations, or specific (Java-based) compilers/tools.

\textbf{Model-to-text transformations (\mtts) \& tools}: Table \ref{tab:pubsources} shows that 64\% of the papers describe \mtts or the generation of code or security infrastructures. 
36\% of the papers do not describe \mtts in details. 
Some mainly used models for verifying or analyzing implemented secure systems, e.g. \textsc{UMLsec} where code/security infrastructure generation is mainly mentioned in future work. 
Comparing the purposes of \mtts, we can see in Fig. \ref{fig:piecharts}c that there are nearly as many \mds papers (34\%) that only generate security infrastructure, such as \textsc{XACML} or security aspects code, as the \mds papers that describe generation of both code and security infrastructure (29\%). 
%A reason might be that some enforcement frameworks for authorisation are platform-independent (e.g. using \textsc{XACML}) and those approaches only focus on generating \textsc{XACML}. 
%Another reason could be that only few approaches support forward or even round-trip engineering for the whole development cycle of secure systems so that both functional code and security infrastructure can be derived. 

% 
%\input{tabresults.tex} 
%\input{tabresults_01.tex}
% Preview source code for paragraph 0

% Add the following just after the closing bracket on this line to specify a position for the table on the page: [h], [t], [b] or [p] - these mean: here, top, bottom and on a separate page, respectively
\begin{table}
\centering % Centers the table on the page, comment out to left-justify
\caption{Results classified by the evaluation criteria}

% Table caption, can be commented out if no caption is required
\setlength{\tabcolsep}{4pt}%
\begin{tabular}{p{3cm}lrr}
Evaluation criteria  &  & \# papers  & \textbf{\%}\tabularnewline
\midrule 
\multirow{5}{3cm}{Security concerns (overlapping)} & Confidentiality  & 45  & \textbf{42}\tabularnewline
 & Integrity  & 29  & \textbf{27}\tabularnewline
 & Availability  & 17  & \textbf{16}\tabularnewline
 & Authenticity  & 26  & \textbf{24}\tabularnewline
 & Authorisation  & 81  & \textbf{75}\tabularnewline
\midrule
\midrule 
\multirow{1}{3cm}{Aspect-Oriented \linebreak{}
Modeling/AOSD} & Yes  & 16  & \textbf{15}\tabularnewline
 & No  & 92  & \textbf{85}\tabularnewline
\midrule 
\multirow{2}{3cm}{Standard models} & Yes(UML/UML profiles)  & 94  & \textbf{87}\tabularnewline
 & Other DSLs  & 14  & \textbf{13}\tabularnewline
\midrule 
\multirow{3}{3cm}{Type of models (overlapping)} & Structural  & 96  & \textbf{89}\tabularnewline
 & Behavioural  & 33  & \textbf{31}\tabularnewline
 & Others  & 14  & \textbf{13}\tabularnewline
\midrule
\midrule 
\multirow{2}{3cm}{Transformations used } & Yes  & 80  & \textbf{74}\tabularnewline
 & No/Unknown  & 28  & \textbf{26}\tabularnewline
\midrule 
\multirow{3}{3cm}{Transformations level } & Endogenous  & 20  & \textbf{19}\tabularnewline
 & Exogenous  & 62  & \textbf{57}\tabularnewline
 & Not Provided  & 26  & \textbf{24}\tabularnewline
\midrule 
\multirow{4}{3cm}{Transformations automation} & Automatic  & 37  & \textbf{34}\tabularnewline
 & Semi-automatic  & 6  & \textbf{6}\tabularnewline
 & Manual  & 4  & \textbf{4}\tabularnewline
 & Not Provided  & 61  & \textbf{56}\tabularnewline
\midrule 
\multirow{2}{3cm}{Standard\linebreak{}
 Transformations } & ATL/QVT  & 20  & \textbf{19}\tabularnewline
 & Others/not mentioned  & 88  & \textbf{81}\tabularnewline
\midrule
\midrule 
\multirow{2}{3cm}{Code generation mentioned} & Yes  & 69  & \textbf{64}\tabularnewline
 & No  & 39  & \textbf{36}\tabularnewline
\midrule 
\multirow{3}{3cm}{Code + Security\linebreak{}
 Infrastructures generated} & Yes  & 31  & \textbf{29}\tabularnewline
 & Only Security Infrastructure  & 37  & \textbf{34}\tabularnewline
 & Not Provided  & 40  & \textbf{37}\tabularnewline
\midrule 
%\multirow{2}{3cm}{Code generation tools} & Xpand/oAW  & \textbf{T}  & \textbf{40}\tabularnewline
% & Others  & \textbf{21}  & \textbf{60}\tabularnewline
%\midrule
\midrule 
\multirow{4}{3cm}{Application\linebreak{}
 Domains} & IS/e-commerce  & 19  & \textbf{18}\tabularnewline
 & Data warehouses  & 20  & \textbf{19}\tabularnewline
 & Smart cards/ embedded systems  & 7  & \textbf{6}\tabularnewline
 & Distributed Systems/SOA  & 34  & \textbf{31}\tabularnewline
 & Others  & 28  & \textbf{26}\tabularnewline
\midrule
\midrule 
\multirow{2}{3cm}{Type of validation} & Controlled experiment  & 2  & \textbf{2}\tabularnewline
 & Industry case studies  & 5  & \textbf{5}\tabularnewline
 & Academic case studies  & 72  & \textbf{67}\tabularnewline
 & Example only & 23  & \textbf{21}\tabularnewline
 & Not Provided  & 6  & \textbf{5}\tabularnewline
\midrule 
 &  &  & \tabularnewline
\end{tabular}\label{tab:pubsources} 
\end{table}

The tools used for code generation are not shown in Table \ref{tab:pubsources} because there are too many different tools. 
Besides Eclipse-based \mtt engines like \textsc{Xpand}\footnote{\url{https://www.eclipse.org/modeling/m2t/?project=xpand}}, there are many cases where ad-hoc self-developed engines (e.g. Java-based tools, parsers, etc.) are used. 
A reason for that could be that many ``ad-hoc'' tools are preferred because of their specific support for a specific security domain.
\textsc{Ark}~\cite{Wada2008}\footnote{extends the code generation engine of the openArchitectureWare framework that was already migrated into Eclipse as \textsc{Xpand}}, for example, transforms an input UML model designed with the proposed UML profile into a skeleton of application code (program code and deployment descriptor). 
More ad-hoc Java-based tools like the one in \cite{Breu2007} generates code (XACML policy files) from the constraints specified in \textsc{SECTET-PL} 
The tool uses Antlr \cite{parr1995antlr}, a compiler program for the syntax analysis of the constraints.

In general, \mmts and \mtts are widely used in \mds to improve the productivity of the development process. 
Most of the primary \mds approaches do mention to leverage \mmts or \mtts by describing transformation rules/intentions. 
However, more than half of the primary \mds approaches did not provide implementation details of \mmts or \mtts. 
Not many primary \mds approaches use standard transformation languages/tools like ATL or QVT but rather ad-hoc tools like Java-based compiler/tools for engineering security into the system. 
With the progress in the maturity of standard \mmt and \mtt tools, they should be leveraged more in the future \mds approaches. 
Most of the \mmts in the selected studies are \emph{exogenous} used for transforming \pim{}s to \psm{}s. 
The main reason is that there are many approaches (e.g. dealing with access control) generating only security infrastructure. % than the approaches generating both code and security infrastructure. 
%Security as a non-functional requirements can be specified in \pim{}s separately from application logic. 
Access control models (\pim{}s) often used to generate \textsc{XACML} configuration files (\psm{}s) for enforcing security policy. %, separately from code. 
Another reason could be the lack of \emph{all-round} approaches for the whole development cycle of secure systems which in the end lead to automatic generation of both code and security infrastructure. 
An \emph{all-round} approach could follow \aom paradigm to fully leverage the automation of \mmts and \mtts for composing, transforming and generating both code and security infrastructure. 
Developing tool chains (based on \mmts and \mtts) to derive from models to implementation code is also an important piece of future work. 
Few complete tool chains to automate (most of) the \mds development process have emerged, but are still rare.

\textbf{Application domains}: Fig. \ref{fig:piecharts}f shows the main application domains that have been secured by \mds approaches. 
In general, these are distributed systems or \soa (31\%), information systems or e-commerce (18\%), data warehouses (19\%), and smart cards/embedded systems (6\%). 
The remaining \mds papers do not clearly state a domain, or could be generically applicable for different application domains, such as \cite{Sanchez2010, Kim2006, Mouheb2010, horcas2014aspect}. 
%\mds approaches for specific application domains seem more mature than general \mds approaches in terms of . 

\textbf{Evaluation methods}: Most of the papers (67\%) describe academic case studies used to evaluate their approaches. 
There are still quite many \mds papers (21\%) which only provide ``running examples"  to illustrate their approaches. 
Few \mds papers show controlled experiments (2\%) and industry case studies (5\%) in the evaluation of their approaches. 
There are very few papers that provide an in-depth evaluation like \cite{Clavel2008}, \cite{Soler2007b}, and \cite{Best2007}. 
Therefore, we suggest that more effort should be put in evaluating \mds approaches, e.g., with empirical studies or benchmarks. 

%The second table \ref{} is the statistic classified according to the main approaches. 
\subsection{Principal \mds Approaches}
\label{sec_princialMDS}

%\twocolumn

Altogether, the synthesised data show that there are currently several \mds approaches that have been proposed, used, and discussed in multiple publications. 
We would like to identify the most influential \mds approaches in terms of numbers of publications and citations. 
%%NOTE: When the letter M is pronounced, it sounds like "em" so since it begins with a vowel sound and we are saying the letters one by one, I would say "an" MDS. 
%In this paper, an \mds approach is called a principal \mds approach if there are at least 7 primary \mds papers in our final set report about this approach. 
In total, five primary \mds approaches, which are called principal \mds approaches, have been identified. 
They are summarised in Table \ref{table:principalMDS}. 
Each has at least $7$ primary \mds papers in our final set. 
The details of each approach, except Secure data warehouses, can be found in \cite{AdvMDS}. 
Here we briefly present each approach, and then compare some key points among them.

\textbf{\textsc{SECTET}} firstly aimed at securing web services by leveraging the Object Constraint Language (\textsc{OCL}) for specifying \textsc{RBAC} \cite{Alam2004}. %\cite{Alam2006b}, \cite{Alam2006a}, \cite{Alam2007a}, 
Based on that, a complete configured security infrastructure (\textsc{XACML} policy files) is generated. 
Later on, the authors proposed a specification language namely \textsc{SECTET-PL} (\textsc{OCL}-based) which is part of the \textsc{SECTET} framework for model-driven security for B2B workflows. 
In this framework, Constraint based \textsc{RBAC} (\textsc{CRBAC}) can be specified and then transformed into low-level web services standard artefacts \cite{Alam2006b}. 
\textsc{SECTET-PL} is also used for modeling restricted (\textsc{RBAC}-based) delegation in Service Oriented Architecture \cite{Alam2006a}. 
Their modeling approach is extended in \cite{Architecture, Policies}. 
\mmt and \mtt are both carried out in a complete model-driven framework \cite{Hafner2006, Breu2007, Breu2008}. 
\textsc{SECTET} mainly addresses \textsc{RBAC} as its security concern and focuses on generating security infrastructure (\textsc{XACML}), not all the source code. 
Recently, \textcite{Memon2012} and also \textcite{Katt2013} propose two pattern refinement approaches based on \textsc{SECTET} framework that allows flexible configurations of SOA security.

\textbf{Secure data warehouses (\textsc{DW}s)} are the motivation for the work of developing \mds techniques for secure database development. 
This \mds approach is very specific for developing secure \textsc{DW}s. 
%(e.g. \cite{Villarroel}, \cite{Soler2007}, \cite{Soler2007a}, \cite{Soler2009}, \cite{Blanco2009}). %\cite{Soler2007b}, 
\textcite{Fernandez-medina2004, Fernandez-medina2004a} extend \textsc{OCL} and \uml for secure database development \cite{Fernandez-Medina2005}. 
Their approach also uses \uml profiles for modelling security enriched \pim{}s as inputs for a model-driven framework to create secure \textsc{DW} solutions \cite{Fernandez-Medina2007, Soler2009}. 
Secure \pim{}s can be transformed to secure \psm{}s by a set of formally defined \textsc{QVT} rules \cite{Soler2007a, Soler2007b, Soler2007}. 
These \psm{}s can then be used for generating code with security properties. 
A similar \mds approach for developing secure XML data warehouses is presented in \cite{Vela2006, Carlos2010, Vela2012, Vela2013}
More recently, the above mentioned techniques for secure \textsc{DW} development are also leveraged in a reverse engineering style to modernise legacy \textsc{DW}s \cite{Blanco2014}.

\providecommand{\tabularnewline}{\\}

%\vspace{-20mm}
\afterpage{
\begin{landscape}
\renewcommand{\arraystretch}{1}
\begin{table}[!p]
\caption{Summary of the Principal \mds Approaches}
\label{table:principalMDS}
{\scriptsize
%\begin{sideways}4
%\begin{tabular}{| p{15mm} | | p{24mm} | p{15mm} |p{5mm}|p{5mm}|p{15mm}|p{15mm}|p{12mm}|p{10mm}|p{12mm}|p{12mm}|p{23mm}|p{12mm}|}
\begin{tabular}{| p{32mm} | | p{20mm} | p{12mm} |p{5mm}|p{5mm}|p{10mm}|p{12mm}|p{7mm}|p{7mm}|p{12mm}|p{13mm}|p{25mm}|p{15mm}|}
\hline 
 & Security &  \multicolumn{4}{c|}{Modeling Approach} & \multicolumn{2}{c|}{MMT} & \multicolumn{2}{c|}{MTT} & Verification 
 %\begin{sideways}
%& Trace-
% \end{sideways} 
& Application  & Validation \tabularnewline

 & Concerns  & \multicolumn{4}{c|}{} & \multicolumn{2}{c|}{} & \multicolumn{2}{c|}{} &  
 %\begin{sideways}
%& ability
% \end{sideways} 
&Domains  &  \tabularnewline

\cline{3-6} \cline{7-8} \cline{9-10} 
 &   & Language & \aom & \soc & Type & Level & Impl & Both & Impl &  &  &  \tabularnewline
\hline 
\hline

\textbf{\textsc{Sectet}} by \textcite{Alam2007a, Alam2006a, Alam2006b, Alam2004, Alam2006, Breu2008, Breu2005, Breu2007, Architecture, Hafner2008, Hafner2005, Hafner2006, Katt2013, Memon2012, Policies}
 
%Security Concerns
& mainly authorisation \linebreak (access control,  delegation),  \linebreak integrity, \linebreak  confidentiality, \linebreak  non-repudiation,

%Modeling Language
& \uml \linebreak profiles
%Paradigm AOM
& \textsf{X}
%Separation of Security
& \checkmark
%Type of Model
& S
%Trafo Level: Endogenous/Exogenous/NA
& Exo
%MMT impl
& \textsc{Qvt}
%Both code + secu infra generated  
& \textsf{X}
%MTT impl
& \textsc{Xpand}  
%Verification
& \textsf{X}
%Application Domains
&
e-government, e-health, \linebreak  e-education, \linebreak  web services, \linebreak  SOA
%Validity Maturity 
& ACS

\tabularnewline
\hline

\textbf{\textsc{SecureDWs}} by  \textcite{Blanco2009a, Blanco2014, Blanco2014a, Blanco, Fernandez-Medina2005, Fernandez-medina2004, Fernandez-medina2004a, Fernandez-Medina2007, Soler2009, Soler2007b, Trujillo2009, Trujillo2009a, Carlos2010, Vela2012, Vela2006, Vela2013, Villarroel2006}

%Security Concerns
& privacy, \linebreak integrity, \linebreak authentication, \linebreak availability, \linebreak non-repudiation, \linebreak auditing, \linebreak access control
%Modeling Language
& \uml \linebreak profiles
%Paradigm
& \textsf{X}
% Separation of Security
& \checkmark 
%Type
 &  S
%Level
& Exo
%MMT
& QVT 
%Traceability  
& \textsf{X} 
%MTT
& MOF, CASE tool
 %Verification
& \textsf{X}
 & 
%Application Domain
web \linebreak applications, databases 
%Validity Maturity 
 & IE, ACS

\tabularnewline
\hline

\textbf{\textsc{SecureMDD}} by \textcite{Moebius2009a, Moebius2009c, Moebius2009b, Moebius2009, Moebius2010, Moebius2012, Borek2012}

%Security Concerns
&  %secrecy, \linebreak integrity, \linebreak confidentiality, 
cryptography (secrecy, \linebreak integrity, \linebreak confidentiality), 
\linebreak application-specific security properties
%Modeling Language
&   \uml \linebreak profiles
%Paradigm
& \textsf{X}
%Separation of Security
&  \checkmark 
 %Verification
& S, B
%Tool Support 
 &  Endo, Exo
  %MMT
& \textsc{Qvt}
%Traceability  
& \checkmark 
 %MTT
& \textsc{Xpand}
%Verification
& \textsc{Kiv} \linebreak theorem prover,\linebreak
   test cases from \uml specifications
 &
%Application Domain
smart card and service \linebreak  applications
%Validity Maturity 
 & IE, ACS, ICS

 \tabularnewline
\hline

\textbf{\textsc{SecureUML}} by \textcite{Lodderstedt2002, Basin2011, Basin2006a, Basin2007, Basin2009, Basin2003, Braga2010, Brucker2006, Clavel2008, Dios2014, basin2014model}
 
%Security Concerns
& access control
%Modeling Language
& \uml \linebreak profiles
%Paradigm
& o
%Separation of Security
& \checkmark
%Verification
& S
%Level
& Endo
%MMT  
& o
%Both gen
& o
%MTT
& ArcStyler, ActionGUI, compiler (self-)
%Verification
& SecureMova model-checker
&
web \linebreak applications
%Validity Maturity 
& IE, ACS, ICS

\tabularnewline
\hline

\textbf{\textsc{UMLsec}} by  \textcite{Jurjens2002, Jorjens2002, Jan2005, Jan2005a, Fox2005, Jan2011, Jurjens2007, Mouratidis2006, Mouratidis2009, Jurjens2006}
 
%Security Concerns
&  confidentiality,\linebreak
integrity,\linebreak authenticity,\linebreak authorisation,\linebreak freshness, \linebreak information flow, \linebreak non-repudiation, \linebreak fair exchange
%Modeing Language
& \uml \linebreak profiles
%Paradigm AOM?
& \textsf{X}
%Separation of Security
& o
%Model Type
& S, B
%Level
& Endo (\cite{Jan2005a, Fox2005})
%Verification
%MMT
& \textsf{X}
%Tool Support 
& \textsf{X}
%MTT
& compiler (self-)
%Verification
& \textsc{aiCall} theorem prover
&
web \linebreak applications, embedded systems, distributed systems
%Validity Maturity 
& IE, ACS, ICS 

\tabularnewline
\hline

 \multicolumn{13}{l}
 {
\input{note1}
 } \tabularnewline
 
  \multicolumn{13}{l}
 {
\input{note2}
 } \tabularnewline

\end{tabular}
%\end{sideways}
}
\end{table}

\end{landscape}
}

\textbf{\textsc{SecureMDD}} %(\cite{Moebius2009a}, \cite{Moebius2009b}, \cite{Moebius2009}, \cite{Moebius2010}, \cite{Moebius2012}) 
is proposed for facilitating the development of smart card applications based on \uml models. 
In \textsc{SecureMDD}, \uml class diagrams are used for modelling static aspects while \uml sequence and activity diagrams are used for modelling dynamic aspects of a system \cite{Moebius2009a}. 
From platform-independent \uml models (\pim{}s) of a system, its formal abstract state machine (ASM) specification and Java Card code are generated. 
The generated abstract state machine specification is used for formally proving the correctness of the generated code regarding the security properties of the system. 
Thus, their \mds approach integrates \mde techniques with semi-formal and formal methods for verification as well as the implementation of security-critical applications \cite{Moebius2009c, Moebius2009, Moebius2010}. 
The authors illustrated that \textsc{SecureMDD} is applicable for the development of large and complex secure Smart Card applications as well \cite{Moebius2012}. 
The main limitations of \textsc{SecureMDD} are its specific application domain and the lack of analysis for consistency between the \uml models and the ASM model.

\textbf{\textsc{SecureUML}} 
%(e.g. \cite{Lodderstedt2002}, \cite{Basin2003}, \cite{Basin2006a}, \cite{Basin2007}, \cite{Brucker2006}) 
is the approach which aims at bridging the gap between security modeling languages and design modeling languages. 
First, \uml and \uml profile are used for modeling application with role-based access control that can lead to generated complete access control infrastructures \cite{Lodderstedt2002}. 
Then, \textcite{Basin2006a} propose a \uml-based language (\uml profiles) with different dialects, which forms modeling languages (such as \textsc{SecureUML} + \textsc{ComponentUML}) for designing secure systems. 
Access control infrastructures for server-based applications can be generated automatically from models. 
Their work mainly focuses on access control constraints based on \textsc{RBAC} in design models. 
Semantics of \textsc{SecureUML} (and \textsc{ComponentUML}) are provided by \textcite{Brucker2006} and \textcite{Basin2007, Basin2009} which enable formal analysis of security-design models. 
Based on this work, Clavel et al. show and discuss their practical experience of applying \textsc{SecureUML} in industrial settings \cite{Clavel2008}. 
Recently, the work on \textsc{SecureUML} has been continued by combining \textsc{SecureUML} + \textsc{ComponentUML} with a language for graphical user interfaces (GUI), namely \textsc{ActionGUI} \cite{basin2014model, Basin2011}. 
These modeling languages with \mmt enable the full generation of security-aware GUIs from models for data-centric applications with access control policies. 
Another recent work by \textcite{Dios2014} makes use of \textsc{ActionGUI} for model driven development of a secure eHealth application. 
The main limitation of \textsc{SecureUML} is its sole focus on access control.

\textbf{\textsc{UMLsec}} 
%(e.g. \cite{Jurjens2002}, \cite{Jan2005}, \cite{Hatebur2011}, \cite{Houmb2003}, \cite{Jurjens2006}) 
is one of the most well-known \uml-based approaches in \mds proposed early by \textcite{Jurjens2002, Jorjens2002}. 
Security requirements, threat scenarios, security concepts, security mechanisms, security primitives can be modeled by using security-related stereotypes (\uml profiles), tags, goal trees. and security constraints. 
Thus, it is possible to formally analyze \textsc{UMLsec} diagrams against security requirements regarding their dynamic behaviors. 
Not like \textsc{SecureUML} only focusing on authorisation (e.g. access control), \textsc{UMLsec} addresses multiple security concerns such as \emph{confidentiality}, \emph{integrity} \cite{Jan2005}. 
Not to a great extent but \aom is also used in the \textsc{UMLsec} approach \cite{Jan2005a}. 
Later on, \textsc{UMLsec} is deployed by \textcite{Best2007} in an industrial context for designing and analyzing designs of distributed information systems. 
On the other hand, relevant tools support for \textsc{UMLsec} are presented in \cite{Jurjens2007}. 
To tackle also social challenges in security, \textsc{UMLsec} was combined with Secure Tropos \cite{mouratidis2007secure} to take on security from requirement engineering phase \cite{Mouratidis2006}. 
This work is then extended and applied to two different industrial case studies \cite{Mouratidis2009}. 
A more recent work related to \textsc{UMLsec} is by \textcite{Jan2011} for incremental security verification for evolving \textsc{UMLsec} models. 
However, \textsc{UMLsec} lacks support for improving productivity of the development process in terms of automated model transformations. 
Even having a view from models to code but the lack of automated transformation(s) from models to implementation code is a miss in \textsc{UMLsec}. 
Other than that, \textsc{UMLsec} could be considered as the most complete and mature \mds approach that deals with multiple security concerns, from very early at the requirement engineering level, with transformations, formal analysis possibility, tools support, industrial case studies. 
%So far, its application domains are for developing secure information systems and secure embedded systems. 

%\cite{Fernandez-medina2004} \cite{Jan2005a} \cite{Villarroel2006},

\textbf{In general}, the most common point among the principal \mds approaches is that they all propose to use \uml profiles in their modeling phase. 
Even though not following truly \aom, defining \uml profiles as \dsl{}s for modeling security concerns still allows these principal \mds approaches to have separation of concerns. 
Except \textsc{SecureUML} which only addresses access control, other approaches are able to touch multiple security concerns. 
Structural models are mainly used in all five approaches. 
\textsc{SecureMDD} and \textsc{UMLsec} have also used behavioral models. 
Exogenous \mmt{}s are defined in \textsc{SECTET} and \textsc{SecureDWs} to transform \pim{}s (\uml models) to \psm{}s. 
\textsc{SecureUML} and \textsc{UMLsec} integrate security into systems specified in \uml using endogenous \mmt{}s. 
\textsc{SecureMDD} combine both kinds of \mmt{}s in their development process. 
Some standard transformation tools are used (e.g. \textsc{QVT} and \textsc{XPAND}) among other self-developed tools (java-based compilers). 
With their formal background, \textsc{SecureMDD}, \textsc{SecureUML} and \textsc{UMLsec} provide some tools for formal verification of security properties. 
These three also have industrial case studies while \textsc{SECTET} and \textsc{SecureDWs} have not. 
Generally, each approach is quite specific to a application domain, e.g. \textsc{SecureDWs} for secure database development, or \textsc{SecureMDD} for secure smart card development.

\subsection{Less common/emerging \mds Approaches}
\label{sec_otherMDSApproaches}

It would not be fair to only discuss about the above-mentioned principal \mds approaches. 
There are other less common or emerging \mds approaches that are also worth to get noticed and analysed. 
We discuss some representative ones here. 
For the full list, readers are referred to Tables \ref{table:otherMDS1} and \ref{table:otherMDS2}. 
The less common or emerging \mds approaches here are simply classified into several groups as follows. 

\textbf{\textsc{Pattern-based} \mds}: Based on domain-independent, time-proven security knowledge and expertise, security patterns can guide security at each stage of the development process. 
Some \mds approaches that leverage security patterns are remarkable. 
%(e.g. \cite{Abramov2012, Abramov2012a, Abramov2012b}, \cite{Bouaziz2013}, \cite{Kim2006, Kim2004, Menzel2010a,Schnjakin2009, moral2014enterprise}). 
\textcite{Abramov2012, Abramov2012a, Abramov2012b} propose an \mds framework for integrating access control policies into database development. 
At the pre-development stage, organisational policies are specified as security patterns. 
Then, the specified security patterns guide the definition and implementation of the security requirements which are defined as part of the data model. 
The database code can be generated automatically after the correct implementation of the security patterns has been verified at the design stage. 
Their approach has been evaluated in a controlled experiment \cite{Abramov2012b}. 
Also using security patterns but at a different level of abstraction, \textcite{Kim2006, Kim2004} develop a pattern-based technique for systematic, model-driven development of secure systems focusing on access control. 
Because this work mainly focuses on the design stage, access control is specified as design pattern. 
\textcite{Bouaziz2013} introduce a security pattern integration process for component-based models. 
With this process, security patterns can be integrated in the whole development process, from UML component modelling until aspect code generation. 
Another pattern-driven approach is proposed by \textcite{Schnjakin2009} for facilitating the configuration of security modules for service-based systems. 
The proposed security advisor enables the transformation from the general security goals, via security patterns at different abstraction level, to concrete security configurations. 
\textcite{Menzel2010a} uses the security configuration patterns to operate the transformation of architecture models annotated with security intentions to security policies. 
The patterns that provide expert knowledge on Web Service security can be specified using a \dsl. 
As using cloud services provided by cloud providers is getting more popular, \textcite{moral2014enterprise} recently propose an enterprise security pattern for securing Software as a Service. 
The security solution provided by the pattern can be driven by making design decisions whilst performing the transformation between the solution models. 
Specifically, from a Computation Independent Model (\textsc{cim}), different \pim{}s can be derived based on different design decisions with security patterns. 
Those \pim{}s are transformed into \psm{}s which are then transformed into Product Dependent Models (\textsc{pdm}s).

\textbf{\mds for \textsc{Security@Runtime}}: Many modern applications such as cloud-based software-as-a-service (SaaS) applications require the dynamic adaptation or even evolution of both security and service at runtime. 
More and more (\mds) approaches have been being proposed in this area. %(e.g. \cite{almorsy2012mdse, Almorsy2013, almorsy2014secdsvl, Elrakaiby2014, Morin2010, Nguyen:2014:TAOSD, Xiao2009}). 
\textcite{almorsy2012mdse} introduce an approach called Model Driven Security Engineering at Runtime (MDSE@R). 
MDSE@R is based on a \uml profile with tool supports for separately specifying base system and security, and then merging those models into a joint system-security model. 
Because security and system models are separated and loosely coupled, they can evolve more easily. 
Security controls are enforced dynamically into the target system at the code level. 
After that, in \cite{Almorsy2013} the same authors leverage the MDSE@R approach for multi-tenant, cloud-hosted SaaS applications. 
This allows dynamically engineering security for multi-tenant SaaS applications at runtime. 
Recently, \textcite{almorsy2014secdsvl} develop a new \dsl called \textsc{SecDVSL} for specifying visually a variety of security concepts like objectives, threats, requirement, architecture, and enforcement controls. 
\textsc{SecDVSL} also allows maintaining traceability among these security concepts. 
Not specifically for SaaS applications but component-based architecture, \textcite{Morin2010} leverage the notion of model@run.time to enable dynamically enforcing role-based access control policies into component-based systems. 
In the follow-up work, \textcite{Nguyen:2014:TAOSD} deal with not only access control policies but also the more complex, but essential, delegation of rights mechanism. 
The propose \mds framework allows dynamically enforcing/weaving access control policies with various delegation features into security-critical systems. 
This is done with a flexibly dynamic adaptation strategy. 
Another runtime-update of security policy-based approach is presented by \textcite{Elrakaiby2014}. 
The introduced \dsl called \textit{Security@Runtime} covers many of the security requirements of modern applications such as authorisation, obligation, and reaction policies. 
\textcite{Xiao2009}'s work is on adaptive and secure multi-agent systems. 
The authors adopting the adaptive agent model to put forward a security-aware model-driven mechanism by using an extension of RBAC model.

\textbf{\mds for \textsc{Secure SOA}}: Many \mds approaches focus on securing service-oriented systems (SOSs). %, e.g. \cite{Gilmore2010, Gallino2012, Hoisl2011, Hoisl2012, Menzel2010a, Menzel2010, Menzel2009, Nakamura2005, Satoh2006, Satoh2007, Wada2008, Wolter2009}. 
%Here we discuss some of them. 
\textcite{Gilmore2010} show how services, service compositions, and non-functional properties can be modeled using their self-developed \uml profile and its extension. 
They address non-functional properties in general where security is considered with performance and reliable messaging. 
The models are the input for the framework VIATRA\footnote{\url{http://www.eclipse.org/viatra/}} to derive deployment mechanisms using \mmt and \mtt. 
\textcite{Wada2008} also address non-functional aspects in SOA with a \mdd framework and tool support. 
Their work is empirically evaluated to show the improvement in the reusability and maintainability of service-oriented applications. 
More specifically to integrate security-related non-functional aspects in the development of services, \textcite{Gallino2012} present their \mds solution using multiple domain-specific models independently addressing security aspects. 
\textcite{Hoisl2011, Hoisl2012} propose an \mds approach based on SoaML for specification and the enforcement of secure object flows in process-driven SOA. 
\cite{Menzel2010, Menzel2009} introduce a security metamodel for SOA. 
This metamodel is the base for their \mds framework that allows modelling of security requirements in system design models. 
Going further than modelling, \textcite{Nakamura2005} propose an \mds tooling framework to generate Web services security configurations. 
In the same line, intermediate model structure is introduced by \textcite{Satoh2006, Satoh2007} to simplify the transformation rules for transforming a security policy written in WebService-SecurityPolicy into platform-specific configuration files. 
%Wolter2009

%\input{tabresults_03.tex}
\providecommand{\tabularnewline}{\\}

%\vspace{-20mm}
\afterpage{
\begin{landscape}
\renewcommand{\arraystretch}{1}
\begin{table}[!p]
\caption{Summary of the less common/emerging \mds Approaches - Part 1}
\label{table:otherMDS1}
{\scriptsize
%\begin{sideways}4
\begin{tabular}{| p{38mm} | | p{19mm} | p{10mm} |p{5mm}|p{5mm}|p{10mm}|p{12mm}|p{7mm}|p{7mm}|p{12mm}|p{10mm}|p{25mm}|p{12mm}|}
\hline 
 & Security &  \multicolumn{4}{c|}{Modeling Approach} & \multicolumn{2}{c|}{MMT} & \multicolumn{2}{c|}{MTT} & Verification 
 %\begin{sideways}
%& Trace-
% \end{sideways} 
& Application  & Validation \tabularnewline

 & Concerns  & \multicolumn{4}{c|}{} & \multicolumn{2}{c|}{} & \multicolumn{2}{c|}{} &  
 %\begin{sideways}
%& ability
% \end{sideways} 
&Domains  &  \tabularnewline

\cline{3-6} \cline{7-8} \cline{9-10} 
 &   & Language & \aom & \soc & Type & Level & Impl & Both & Impl &  &  &  \tabularnewline
\hline 
\hline

\textbf{\textsc{Pattern-Based}} by \textcite{Abramov2012, Abramov2012a, Abramov2012b}
 
%Security Concerns
& access control
%Modelling Language
& \uml
%Paradigm AOM
& \textsf{X}
%Separation of Security
& \checkmark
%Type of Model
& S
%Trafo Level: Endogenous/Exogenous/NA
& Exo
%MMT impl
& ATL
%Both code + secu infra generated  
& \checkmark
%MTT impl
& SMT (self-)
%Verification
& \checkmark
%Application Domains
&
database
%Validity Maturity 
& ACS; CE

\tabularnewline
\hline

\textbf{\textsc{Pattern-Based}} by \textcite{Bouaziz2013}
 
%Security Concerns
& access control
%Modeling Language
& \uml
%Paradigm AOM
& \textsf{X}
%Separation of Security
& \checkmark
%Type of Model
& S
%Trafo Level: Endogenous/Exogenous/NA
& Endo
%MMT impl
& ATL
%Both code + secu infra generated  
& \checkmark
%MTT impl
& NP
%Verification
& \textsf{X}
%Application Domains
&
component-based architecture
%Validity Maturity 
& ACS

\tabularnewline
\hline

\textbf{\textsc{Pattern-Based}} by \textcite{Kim2006, Kim2004}
 
%Security Concerns
& access control
%Modeling Language
& \uml
%Paradigm AOM
& \checkmark
%Separation of Security
& \checkmark
%Type of Model
& S, B
%Trafo Level: Endogenous/Exogenous/NA
& Endo, Exo
%MMT impl
& NP
%generated  
& \textsf{X}
%MTT impl
& \textsf{X}
%Verification
& \checkmark
%Application Domains
&
NR
%Validity Maturity 
& ACS

\tabularnewline
\hline

\textbf{\textsc{Pattern-Based}} by \textcite{Schnjakin2009}
 %Security Concerns
& integrity, \linebreak  confidentiality, \linebreak  authentication,  \linebreak authorisation
%Modeling Language
& BPMN
%Paradigm AOM
& \textsf{X}
%Separation of Security
& \checkmark
%Type of Model
& O
%Trafo Level: Endogenous/Exogenous/NA
& NP
%MMT impl
& \textsf{X}
%Both code + secu infra generated  
& \textsf{X}
%MTT impl
& NP
%Verification
&  \textsf{X}
%Application Domains
&
service-oriented architectures
%Validity Maturity 
& IE

\tabularnewline
\hline

\textbf{\textsc{Pattern-Based}} by \textcite{moral2014enterprise}
 
%Security Concerns
& integrity, \linebreak  confidentiality,  \linebreak  availability,  \linebreak  authentication,  \linebreak authorisation
%Modeling Language
& \dsl
%Paradigm AOM
& \textsf{X}
%Separation of Security
& \checkmark
%Type of Model
& O
%Trafo Level: Endogenous/Exogenous/NA
& Exo
%MMT impl
& \textsf{X}
%Both code + secu infra generated  
& \textsf{X}
%MTT impl
& NP
%Verification
&  \textsf{X}
%Application Domains
&
secure cloud computing
%Validity Maturity 
& IE

\tabularnewline
\hline

\textbf{\textsc{Sec@runtime}} by \textcite{almorsy2012mdse, Almorsy2013, almorsy2014secdsvl}
%Security Concerns
& integrity, \linebreak  confidentiality, \linebreak  availability, \linebreak  authentication,  \linebreak authorisation
%Modeling Language
& \uml
%Paradigm
& \checkmark
% Separation of Security
& \checkmark
%Type
 &  S, B
%Level
& Endo
%MMT
& NP
% Both gen  
& o
%MTT Impl
& NP
 %Verification
& testing
 & 
%Application Domain
cloud-based applications
%Validity Maturity 
 &   ACS, CE

\tabularnewline
\hline

\textbf{\textsc{Sec@runtime}} by \textcite{Elrakaiby2014}
%Security Concerns
& authorisation
%Modeling Language
& \uml, \dsl
%Paradigm
& \checkmark
% Separation of Security
& \checkmark
%Type
 &  S, \dsm
%Level
& Exo
%MMT
& NP
%Both gen  
& \textsf{X}
%MTT
& NP
 %Verification
& \textsf{X}
 & 
%Application Domain
 NR
%Validity Maturity 
 &   ACS

\tabularnewline
\hline

\textbf{\textsc{Sec@runtime}} by \textcite{Morin2010}
%Security Concerns
& authorisation
%Modeling Language
& \dsl
%Paradigm
& \textsf{X}
% Separation of Security
& \checkmark
%Type
 & \dsm
%Level
& Exo
%MMT
& Kermeta
%Traceability  
& o
%MTT
& Kermeta
 %Verification
& \textsf{X}
 & 
%Application Domain
component-based architecture
%Validity Maturity 
 &  ACS

\tabularnewline
\hline

\textbf{\textsc{Sec@runtime}} by \textcite{Nguyen:2014:TAOSD}
%Security Concerns
& authorisation (access control, delegation)
%Modeling Language
& \dsl
% Separation of Security
& \textsf{X}
%Type
 &  \checkmark
 %Paradigm
& \dsm
%Level
& Exo
%MMT
& Kermeta
%Traceability  
& o
%MTT
& Kermeta
 %Verification
& \textsf{X}
 & 
%Application Domain
component-based architecture
%Validity Maturity 
 &  ACS

\tabularnewline
\hline

\textbf{\textsc{Sec@runtime}} by \textcite{Xiao2009}
%Security Concerns
& authorisation
%Modeling Language
& \dsl
%Paradigm
& \dsm
% Separation of Security
& \textsf{X}
%Type
 &  \checkmark
%Level
& Exo
%MMT
& NP
%Traceability  
& \checkmark
%MTT
& JADE(self-)
 %Verification
& \textsf{X}
 & 
%Application Domain
NR
%Validity Maturity 
 &  ACS

\tabularnewline
\hline

\textbf{\textsc{SecureSOA}} by \textcite{Gallino2012}
%Security Concerns
& authorisation
%Modeling Language
& \uml
%Paradigm
& \textsf{X}
% Separation of Security
& \checkmark
%Type
 &  S
%Level
& Exo
%MMT
& NP
%Traceability  
& o
%MTT
& NP
 %Verification
& \textsf{X}
 & 
%Application Domain
SOA
%Validity Maturity 
 &   ACS

\tabularnewline
\hline

\textbf{\textsc{SecureSOA}} by \textcite{Gilmore2010}
%Security Concerns
& non-functional aspects
%Modeling Language
& \uml  \linebreak profiles
%Paradigm
& \textsf{X}
% Separation of Security
& o
%Type
 &  S, B, O
%Level
& Exo
%MMT
& VIATRA
%Traceability  
& \checkmark
%MTT
& VIATRA2
 %Verification
& \checkmark
 & 
%Application Domain
distributed systems
%Validity Maturity 
 &   ACS

\tabularnewline
\hline

\textbf{\textsc{SecureSOA}} by \textcite{Hoisl2011, Hoisl2012}
%Security Concerns
& confidentiality, integrity
%Modeling Language
& \uml
%Paradigm
& \textsf{X}
% Separation of Security
& \checkmark
%Type
 &  S, B
%Level
& Exo
%MMT
& NP
%Traceability  
& NP
%MTT
& NP
 %Verification
& \textsf{X}
 & 
%Application Domain
SOA
%Validity Maturity 
 &   ACS

\tabularnewline
\hline

\textbf{\textsc{SecureSOA}} by \textcite{Menzel2010a, Menzel2010, Menzel2009}
%Security Concerns
& confidentiality, integrity, authentication, authorisation
%Modeling Language
& \dsl
%Paradigm
&  \textsf{X}
% Separation of Security
& \checkmark
%Type
 &  \dsm
%Level
& Exo
%MMT
& NP
%Traceability  
& o
%MTT
& NP
 %Verification
& \textsf{X}
 & 
%Application Domain
SOA
%Validity Maturity 
 &   ACS

\tabularnewline
\hline

\textbf{\textsc{SecureSOA}} by \textcite{Nakamura2005}
%Security Concerns
& confidentiality, integrity, availability, authentication
%Modeling Language
& \uml profile
%Paradigm
& \textsf{X}
% Separation of Security
& \checkmark
%Type
 &  S
%Level
& Exo
%MMT
& NP
%Traceability  
& o
%MTT
& NP
 %Verification
& \textsf{X}
 & 
%Application Domain
SOA
%Validity Maturity 
 &   IE

\tabularnewline
\hline

\textbf{\textsc{SecureSOA}} by \textcite{Satoh2006, Satoh2007}
%Security Concerns
& authentication
%Modeling Language
& \dsl
%Paradigm
& \textsf{X}
% Separation of Security
& \checkmark
%Type
 &  \dsm
%Level
& Endo
%MMT
& NP
%Traceability  
& o
%MTT
& NP
 %Verification
& \textsf{X}
 & 
%Application Domain
SOA
%Validity Maturity 
 &   IE

\tabularnewline
\hline

\textbf{\textsc{SecureSOA}} by \textcite{Wada2008}
%Security Concerns
& confidentiality, integrity, authorisation
%Modeling Language
& \uml profile
%Paradigm
& \textsf{X}
% Separation of Security
& \checkmark
%Type
 &  S
%Level
& Exo
%MMT
& ark
%Traceability  
& \checkmark
%MTT
& ark (self-)
 %Verification
& \textsf{X}
 & 
%Application Domain
SOA
%Validity Maturity 
 &  CE%, Empirical Study

\tabularnewline
\hline

\textbf{\textsc{SecureSOA}} by \textcite{Wolter2009}
%Security Concerns
& integrity, \linebreak  confidentiality, \linebreak  availability, \linebreak  authentication,  \linebreak authorisation
%Modeling Language
& \dsl
%Paradigm
& \textsf{X}
% Separation of Security
& \checkmark
%Type
 &  \dsm
%Level
& Exo
%MMT
& NP
%Traceability  
& o
%MTT
& NP
 %Verification
& \textsf{X}
 & 
%Application Domain
SOA
%Validity Maturity 
 &   IE

\tabularnewline
\hline

 \multicolumn{13}{l}
 {
\input{note1}
 } \tabularnewline
 
  \multicolumn{13}{l}
 {
\input{note2}
 } \tabularnewline
 
\end{tabular}
%\end{sideways}
}
\end{table}

\end{landscape}
}

\providecommand{\tabularnewline}{\\}

%\vspace{-20mm}
\afterpage{
\begin{landscape}
\renewcommand{\arraystretch}{1}
\begin{table}[!p]
\caption{Summary of the less common/emerging \mds Approaches - Part 2}
\label{table:otherMDS2}
{\scriptsize
%\begin{sideways}4
\begin{tabular}{| p{32mm} | | p{20mm} | p{12mm} |p{5mm}|p{5mm}|p{10mm}|p{12mm}|p{7mm}|p{7mm}|p{12mm}|p{10mm}|p{25mm}|p{15mm}|}
\hline 
 & Security &  \multicolumn{4}{c|}{Modeling Approach} & \multicolumn{2}{c|}{MMT} & \multicolumn{2}{c|}{MTT} & Verification 
 %\begin{sideways}
%& Trace-
% \end{sideways} 
& Application  & Validation \tabularnewline

 & Concerns  & \multicolumn{4}{c|}{} & \multicolumn{2}{c|}{} & \multicolumn{2}{c|}{} &  
 %\begin{sideways}
%& ability
% \end{sideways} 
&Domains  &  \tabularnewline

\cline{3-6} \cline{7-8} \cline{9-10} 
 &   & Language & \aom & \soc & Type & Level & Impl & Both & Impl &  &  &  \tabularnewline
\hline 
\hline

\textbf{\textsc{AOMsec}} by \textcite{Georg2009}
%Security Concerns
& integrity, \linebreak  confidentiality, \linebreak  availability, \linebreak  authentication,  \linebreak authorisation
%Modeling Language
& \uml
%Paradigm
& \checkmark
% Separation of Security
& \checkmark
%Type
 &  S, B
%Level
& Endo
%MMT
& NP
%Traceability  
& NP
%MTT
& NP
 %Verification
& \checkmark
 & 
%Application Domain
NR
%Validity Maturity 
 &  IE

\tabularnewline
\hline

\textbf{\textsc{AOMsec}} by \textcite{Mouheb2010, Mouheb2009a}
%Security Concerns
& confidentiality, authorisation
%Modeling Language
& \uml
%Paradigm
& \checkmark
% Separation of Security
& \checkmark
%Type
 &  S, B, O
%Level
& Endo
%MMT
& QVT
%Traceability  
& \checkmark
%MTT
& RSA
 %Verification
& \textsf{X}
 & 
%Application Domain
NR
%Validity Maturity 
 &  ACS

\tabularnewline
\hline

\textbf{\textsc{AOMsec}} by \textcite{Ray2004575}
%Security Concerns
& authorisation (AC)
%Modeling Language
& \uml
%Paradigm
& \checkmark
% Separation of Security
& \checkmark
%Type
 &  S, O
%Level
& Endo
%MMT
& NP
%Traceability  
& NP
%MTT
& NP
 %Verification
& \textsf{X}
 & 
%Application Domain
NR
%Validity Maturity 
 &  IE

\tabularnewline
\hline

\textbf{\textsc{AOMsec}} by \textcite{Sanchez2010}
%Security Concerns
& confidentiality, integrity, authorisation
%Modeling Language
& \uml profile
%Paradigm
& \checkmark
% Separation of Security
& \checkmark
%Type
 &  S, B
%Level
& Exo
%MMT
& QVT
%Traceability  
& o
%MTT
& NP
 %Verification
& \textsf{X}
 & 
%Application Domain
NR
%Validity Maturity 
 &  ACS

\tabularnewline
\hline

\textbf{\textsc{AOMsec}} by \textcite{Zhu2009}
%Security Concerns
& confidentiality, integrity, availability
%Modeling Language
& \uml profile
%Paradigm
& \checkmark
% Separation of Security
& \checkmark
%Type
 &  S, B
%Level
& Exo
%MMT
& NP
%Traceability  
& o
%MTT
& aspect code gen (self-)
 %Verification
& \checkmark
 & 
%Application Domain
NR
%Validity Maturity 
 &  ICS

\tabularnewline
\hline

\textbf{\textsc{Access Control oriented}} by \textcite{Ahn2007}
%Security Concerns
& authorisation
%Modeling Language
& \uml
%Paradigm
& \textsf{X}
% Separation of Security
& \textsf{X}
%Type
 &  S, O
%Level
& NP
%MMT
& NP
%Traceability  
& o
%MTT
& Octopus + Dresden OCL toolkit
 %Verification
& \textsf{X}
 & 
%Application Domain
NR
%Validity Maturity 
 &  CE

\tabularnewline
\hline

\textbf{\textsc{Access Control oriented}} by \textcite{Burt2003}
%Security Concerns
& authorisation
%Modeling Language
& \dsl
%Paradigm
& \textsf{X}
% Separation of Security
&  \checkmark
%Type
 &  \dsm
%Level
& Exo
%MMT
& NP
%Traceability  
& NP
%MTT
& NP
 %Verification
& \textsf{X}
 & 
%Application Domain
NR
%Validity Maturity 
 &  IE

\tabularnewline
\hline

\textbf{\textsc{Access Control oriented}} by \textcite{Fink2006}
%Security Concerns
& authorisation
%Modeling Language
& \dsl
%Paradigm
&  \textsf{X}
% Separation of Security
& \checkmark
%Type
 &  \dsm
%Level
& Exo
%MMT
& Graph Transformation
%Traceability  
& NP
%MTT
& NP
 %Verification
&  \textsf{X}
 & 
%Application Domain
NR
%Validity Maturity 
 &  ACS

\tabularnewline
\hline

\textbf{\textsc{Access Control oriented}} by \textcite{Kim2011}
%Security Concerns
& authorisation (AC)
%Modeling Language
& \uml
%Paradigm
& \checkmark
% Separation of Security
& \checkmark
%Type
 &  S, B
%Level
& Endo
%MMT
& IBM RSA
%Traceability  
& \checkmark
%MTT
& IBM RSA
 %Verification
& o
 & 
%Application Domain
NR
%Validity Maturity 
 &  ACS

\tabularnewline
\hline

\textbf{\textsc{Access Control oriented}} by \textcite{Mouelhi2008}
%Security Concerns
& authorisation
%Modeling Language
& \dsl
%Paradigm
&  \textsf{X}
% Separation of Security
& \checkmark
%Type
 &  \dsm
%Level
& Exo
%MMT
& NP
%Traceability  
& o
%MTT
& NP
 %Verification
& \checkmark (testing)
 & 
%Application Domain
NR
%Validity Maturity 
 &  ACS

\tabularnewline
\hline

\textbf{\textsc{Access Control oriented}} by \textcite{kaddani2014towards}
%Security Concerns
& authorisation (OrBAC)
%Modeling Language
& \uml
%Paradigm
&  \textsf{X}
% Separation of Security
& \checkmark
%Type
 &  S
%Level
& Exo
%MMT
& NP
%code gen  
& o
%MTT
& NP
 %Verification
& \textsf{X}
 & 
%Application Domain
electrical grid
%Validity Maturity 
 &  IE 

\tabularnewline
\hline

\textbf{\textsc{Access Control oriented}} by \textcite{Pavlich-Mariscal2010}
%Security Concerns
& authorisation, authentication
%Modeling Language
& \uml profile
%Paradigm
&  \textsf{X}
% Separation of Security
& \checkmark
%Type
 &  S
%Level
& Endo
%MMT
& NP
%Traceability  
& \checkmark
%MTT
& self-
 %Verification
& o
 & 
%Application Domain
NR
%Validity Maturity 
 &  ACS

\tabularnewline
\hline

\textbf{\textsc{Access Control oriented}} by \textcite{Sohr2008}
%Security Concerns
& authorisation
%Modeling Language
& \uml
%Paradigm
& \textsf{X}
% Separation of Security
& \checkmark
%Type
 &  S, B, O
%Level
& NP
%MMT
& NP
%Traceability  
& NP
%MTT
& NP
 %Verification
& \textsf{X}
 & 
%Application Domain
distributed systems
%Validity Maturity 
 &   ACS

\tabularnewline
\hline

\textbf{\textsc{Access Control oriented}} by \textcite{schefer2014model}
%Security Concerns
& authorisation
%Modeling Language
& \uml profile
%Paradigm
& \textsf{X}
% Separation of Security
& \checkmark
%Type
 &  S, B, O
%Level
& Exo
%MMT
& NP
%Traceability  
& NP
%MTT
& NP
 %Verification
& \textsf{X}
 & 
%Application Domain
NR
%Validity Maturity 
 &   ACS

\tabularnewline
\hline

\textbf{\textsc{Access Control oriented}} by \textcite{bertolino2014toolchain}
%Security Concerns
& authorisation
%Modeling Language
& \dsl
%Paradigm
& \textsf{X}
% Separation of Security
& \checkmark
%Type
 &  S, B
%Level
& Exo
%MMT
& Kermeta
%Traceability  
& o
%MTT
& NP
 %Verification
& model-based testing
 & 
%Application Domain
NR
%Validity Maturity 
 &   ACS

\tabularnewline
\hline

\textbf{\textsc{Usage Control}} by \textcite{Neisse2013}
%Security Concerns
& authorisation (UCON)
%Modeling Language
& \dsl
%Paradigm
& \textsf{X}
% Separation of Security
& \checkmark
%Type
 &  \dsm, S, B, O
%Level
& Exo
%MMT
& Java-based tool
%Traceability  
& o
%MTT
& Java-based tool (self-)
 %Verification
& \textsf{X}
 & 
%Application Domain
NR
%Validity Maturity 
 &  ACS

\tabularnewline
\hline

\textbf{\textsc{ModelSec}} by \textcite{Sanchez2009}
%Security Concerns
& integrity, \linebreak  confidentiality, \linebreak  availability, \linebreak  authentication,  \linebreak authorisation
%Modeling Language
& \dsl (SecML)
%Paradigm
& \textsf{X}
% Separation of Security
& \checkmark
%Type
 &  \dsm
%Level
& Exo
%MMT
& RubyTL
%Traceability  
& o
%MTT
& MOFScript
 %Verification
& \textsf{X}
 & 
%Application Domain
NR
%Validity Maturity 
 &  ACS

\tabularnewline
\hline 

\textbf{\textsc{Secure Web Apps}} by \textcite{busch2014modeling}
%Security Concerns
& integrity, \linebreak  confidentiality, \linebreak  availability, \linebreak  authentication,  \linebreak authorisation
%Modeling Language
& \uml profile
%Paradigm
& \textsf{X}
% Separation of Security
& \checkmark
%Type
 &  \dsm
%Level
& Exo
%MMT
& NP
%Traceability  
& o
%MTT
& XPand
 %Verification
& testing possible
 & 
%Application Domain
web applications
%Validity Maturity 
 &  ACS

\tabularnewline
\hline

\textbf{\textsc{SecEmbedded}} by \textcite{Eby2007}
%Security Concerns
& confidentiality, availability
%Modeling Language
& \dsl
%Paradigm
& \textsf{X}
% Separation of Security
& \checkmark
%Type
 &  \dsm
%Level
& Exo
%MMT
& NP
%Traceability  
& NP
%MTT
& NP
 %Verification
& \textsf{X}
 & 
%Application Domain
embedded systems
%Validity Maturity 
 &   ACS

\tabularnewline
\hline

 \multicolumn{13}{l}
 {
\input{note1}
 } \tabularnewline
 
  \multicolumn{13}{l}
 {
\input{note2}
 } \tabularnewline
 
\end{tabular}
%\end{sideways}
}
\end{table}

\end{landscape}
}

\textbf{\textsc{Aspect-Oriented Modelling} in \mds}: \aom techniques would be ideal for \mds with fully separation of concerns support. 
With \aom, security concerns can be modelled separately, and then automatically composed into primary models. 
All of the reviewed \mds approaches in this category except \cite{Ray2004575, Zhu2009} tackle multiple security concerns. 
%The different \mds approaches proposed by \textcite{Georg2009, Mouheb2010, Mouheb2009a, Sanchez2010, horcas2014aspect} 
These approaches aim at dealing with multiple security concerns as one would expect from any \aom approach. 
\textcite{Georg2009} propose a methodology that allows not only security mechanisms but also attacks to be modelled as aspect models. 
The attacks models can be composed with the primary model of the application to obtain the misuse model. 
The authors then use the Alloy Analyser \footnote{\url{http://alloy.mit.edu}} to reason about the misuse model. 
If the misuse model shows that the application is compromised, some security mechanism must be incorporated into the application. 
The Alloy Analyser is used again to verify that the secured application model is now resilient to the attack. 
\textcite{Mouheb2010, Mouheb2009a} develop a \uml profile that allows specifying security mechanisms as aspect models. 
The aspect models often go together with their integration specification. 
Their approach allows security aspects to be woven automatically into UML design models (class diagrams, state machine diagrams, sequence diagrams, and activity diagrams) \cite{Mouheb2010}. 
In \cite{Mouheb2009a}, the authors present a full security hardening approach, from design to implementation. 
Not only restricted to security aspects, \textcite{Sanchez2010} propose a \mdd approach for all early aspects, including security. 
The difference with other approaches is that they focus on aspect-oriented requirements specifications (models). 
These aspect-oriented requirements models are then automatically transformed into aspect-oriented architecture models. 
Not dealing with multiple security concerns, \textcite{Ray2004575} introduce an \aom approach for addressing access control. 
Specifically, RBAC aspects can be modeled using parameterised \uml models as patterns. 
This allows uniformly incorporate pervasive access control functionality into a design. 
The woven model can be analysed to check the correctness of incorporation. 
\textcite{Zhu2009} propose a model-based aspect-oriented framework for building intrusion-aware software systems. 
There, attack scenarios and intrusion detection aspects are modelled using an aspect-oriented \uml profile. 
The intrusion detection aspect models are used to automatically generate aspect-oriented codes. 
The aspect-oriented codes are woven into the target systems using an aspect weaver to obtain the intrusion-aware software system. 
Recently, \textcite{horcas2014aspect} propose a hybrid \aosd and \mde approach for automatically weaving a customised security model into the base application model. 
By using the Common Variability Language (\textsc{CVL}) and \textsc{atl}, different security concerns can be woven into the base application in an aspect-oriented way, according to weaving patterns. 
However, inter-security concern relations have not been taken into account.

\textbf{\mds for \textsc{Access Control}}: Section \ref{sec_perCriterion} shows that access control problem got the most attention from the \mds community. 
We discuss here some representative \mds approaches that specifically address access control. % (\cite{Ahn2007, Fink2006, Kim2011, Mouelhi2008, Pavlich-Mariscal2010, schefer2014model, bertolino2014toolchain}). 
\textcite{Ahn2007} propose a framework for representing security model, specifying and validating security policy, and automatically generating security enforcement codes. 
This framework leverages the \mdd approach together with a systematic tool to build secure systems. 
Also presenting a \mdd approach for access control, \textcite{Fink2006} aim at developing access control policies for distributed systems using \mof and \uml profiles. 
However, this approach does not work well with module-based system like systems based on SOAP \footnote{\url{http://www.w3.org/TR/soap/}}. 
\textcite{Kim2011} present a feature-based approach that enables systematic configuration of RBAC features for developing customisable access control-based enterprise systems. 
Feature modelling is used for effectively capturing the variabilities of the RBAC. 
UML models are used for specifying the static and behavioural properties of RBAC features. 
The composition method in their approach is used for building RBAC configuration, which also serves as a verification point for correctness of composition. 
Aiming at a full design-to-testing \mdd process, \textcite{Mouelhi2008} introduce a generic access control metamodel. 
The generic access control policy model specified by the metamodel is automatically transformed into security policy for the XACML platform, and integrated in the target application using aspect-oriented programming. 
Model-based mutation testing makes the access control enforcement quantitatively testable. 
\textcite{Pavlich-Mariscal2010} propose a \md framework with a set of composable access control features that can be tightly integrated into the \uml. 
At the code level, access control is map to the policy code which realises access control diagrams and features, and the enforcement code, to restrict access to methods based on information of the policy code. 
The degree of traceability of mappings is assessed. 
Recently, \textcite{schefer2014model} propose a full \mdd approach for specifying and enforcing break-glass policies in process-aware information systems. 
By tackling a complex security exception handling mechanism like break-glass policies with \mds, this work shows developing \dsl{}s for specific security concerns are a good way to capture well the semantics of these concerns. 
Based on that, a typical \mdd process can be developed for derive security from specification to enforcement with tools support. 
\textcite{bertolino2014toolchain} even go further in terms of tools support by providing a toolchain for designing, generating. and testing access control policies. 
This toolchain is the result of integrating specific tools for specific stages of the development cycle that have been developed in a collaborative research network. 
The research around \umlsec has also resulted in various tools support but not yet systematically formed a tool chain.

\textbf{Miscellaneous}: \textcite{Neisse2013} present one of few \mds approaches about usage control, the next generation of access control. 
Consisting of authorisations and obligations, high-level usage control policies are specified considering an abstract system model and automatically refined with the help of policy refinement rules to implementation-level policies. 
The work by \textcite{Elrakaiby2014} mentioned above can also be categorised as usage control. 
In the domain of securing embedded systems, the approach we reviewed is by \textcite{Eby2007}. 
The authors propose a framework to incorporate security modelling into embedded system design. 
Their security analysis tool is capable of analysing the flow of data objects through a system and identifying points that are vulnerable to attack. 
Not restricted to a particular application domain, \textsc{ModelSec} by \textcite{Sanchez2009} can deal with multiple security concerns in an integrated fashion, including privacy, integrity, access control, authentication, availability, non-repudiation, and auditing. 
\textsc{ModelSec} supports defining and managing security requirements by building security requirements models for an application from which operational security models can then be generated.
Recently, \textcite{busch2014modeling} present an \mds approach specific for securing web applications, tackling multiple security concerns. 
The graphical, \uml{}-based Web Engineering (\textsc{UWE}) language is extended for specifying security concerns in web applications. 
Moreover, the approach is mapped to an iterative development cycle from requirement specification to testing and deployment with tools support.

\subsection{Trend analysis of \mds approaches}
\label{sec_trend}

\begin{figure}
\center
\includegraphics[width=0.5\textwidth]{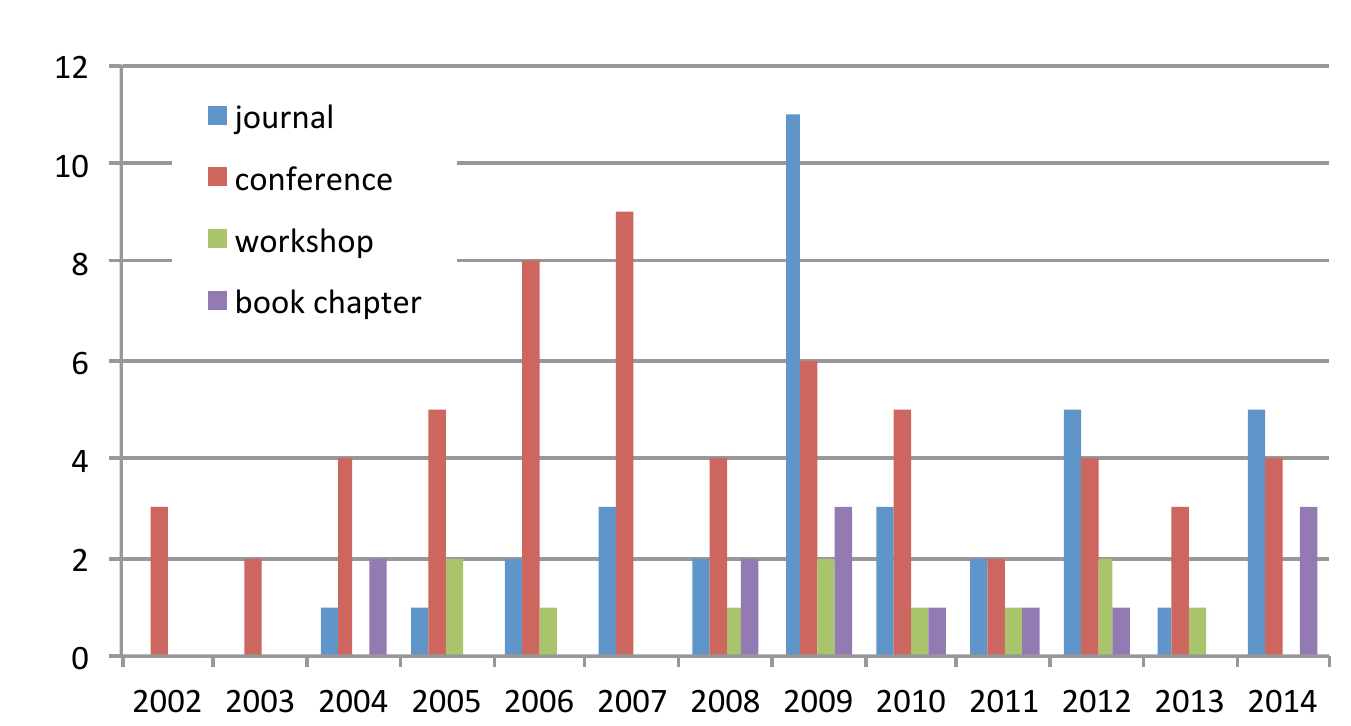}
\caption{Trend of \mds publication}%
\label{fig:venue_type}
\end{figure}

%\begin{figure*}
%\center
%\includegraphics[scale=0.7]{Figs/trends_Phu}
%\caption{TODO}%
%\label{fig:piecbahbaharts}
%\end{figure*}

In terms of publication, we can see in Fig. \ref{fig:venue_type} there was a peak time for primary \mds publications in 2009. 
As we mentioned, the primary \mds approaches were first introduced from 2002. 
From 2002-2008, more primary \mds papers were published at conferences than journals. 
The number of primary \mds papers published at conferences were going up until 2007. 
In 2008, the number of primary \mds papers published at conferences decreased. 
One of the reasons could be primary \mds papers were under submission to journals. 
In 2009, there was a peak number of primary \mds papers published in journals. 
After the peak in 2009, the trend of primary \mds publications looks more stable for the period 2010-2014. 
From 2010 to 2014, less primary \mds papers were published than the previous 5-year period (2005-2009). 
However, the trend of publishing primary \mds papers in the period 2010-2014 seems more stable.

\begin{figure}
\center
\includegraphics[width=0.5\textwidth]{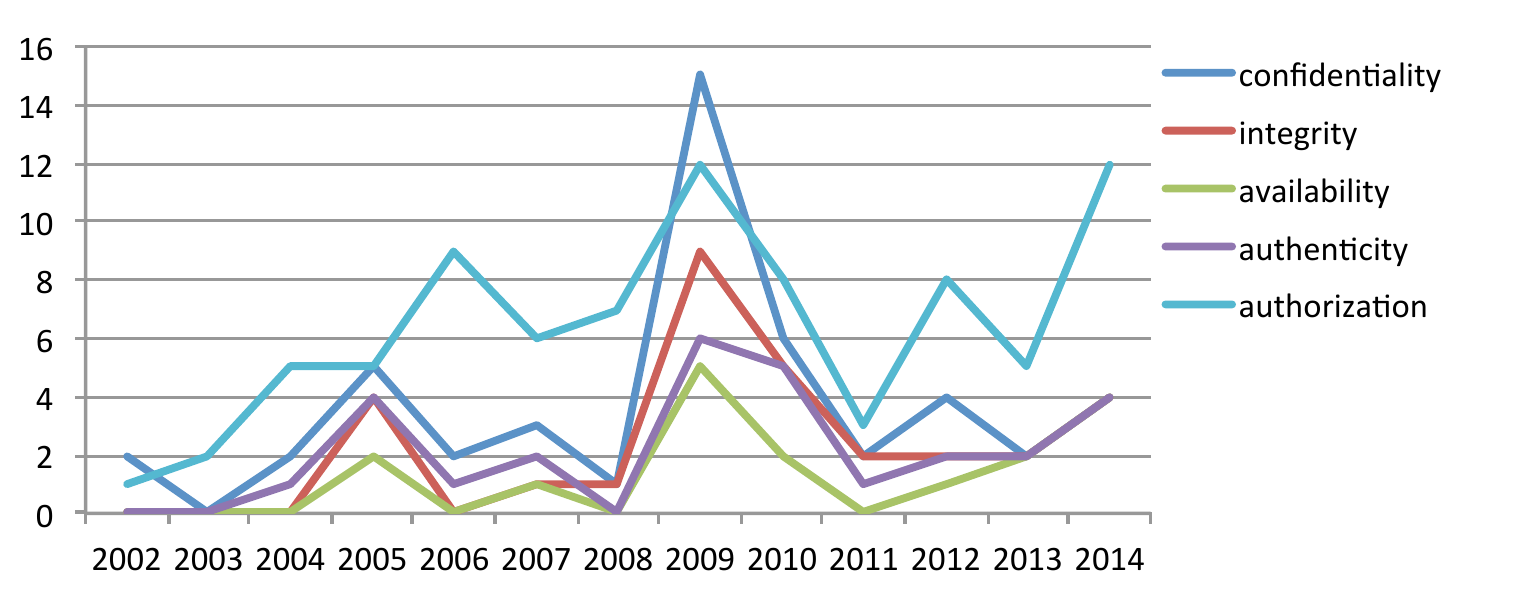}
\caption{Trend of security concerns addressed by \mds studies}%
\label{fig:characteristics_trend}
\end{figure}

Similarly to the trend of publications, the trend of how security concerns have been addressed also has a peak time in 2009. 
Fig. \ref{fig:characteristics_trend} shows that, nearly all the time reviewed, authorisation is the concern that has been addressed the most. 
Only in 2009, confidentiality was tackled by more primary \mds papers than authorisation. 
The other concerns were always less focused than authorisation and confidentiality all the time reviewed. 
Until 2014, authorisation looks like still being addressed the most by the \mds research community. 
\mds researchers should pay more attention to the less tackled security concerns, and should aim at a solution addressing multiple security concerns simultaneously.

\begin{figure}
\center
\includegraphics[width=0.5\textwidth]{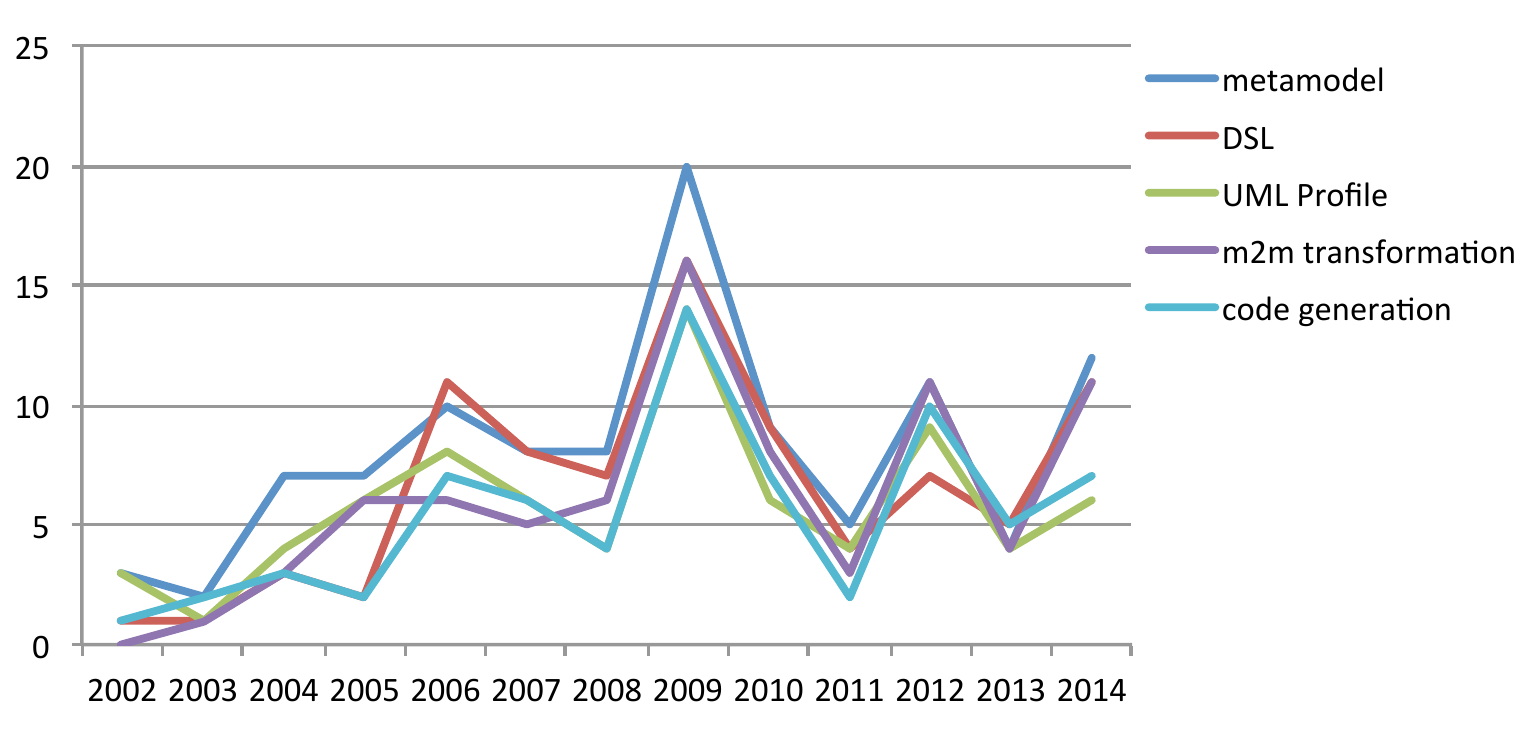}
\caption{Trend of \mde artefacts leveraged by \mds studies}%
\label{fig:mde_trend}
\end{figure}

The trends of how \mde artefacts leveraged in the primary \mds approaches look well coupled with the number of primary \mds publications. 
The line of each artefact is very close to the others (see Fig. \ref{fig:mde_trend}). 
This means that most primary \mds approaches did leverage the key artefacts of \mde in secure systems development. 
It is easily understandable that as long as we clearly define how an approach can be considered an \mds approach, 
most of the key \mde artefacts have to be leveraged in an \mds approach. 
This trend should hold in the future as well.

\begin{figure*}[htbp]
\center
\subfloat[journals]{\includegraphics[width=0.40\textwidth]{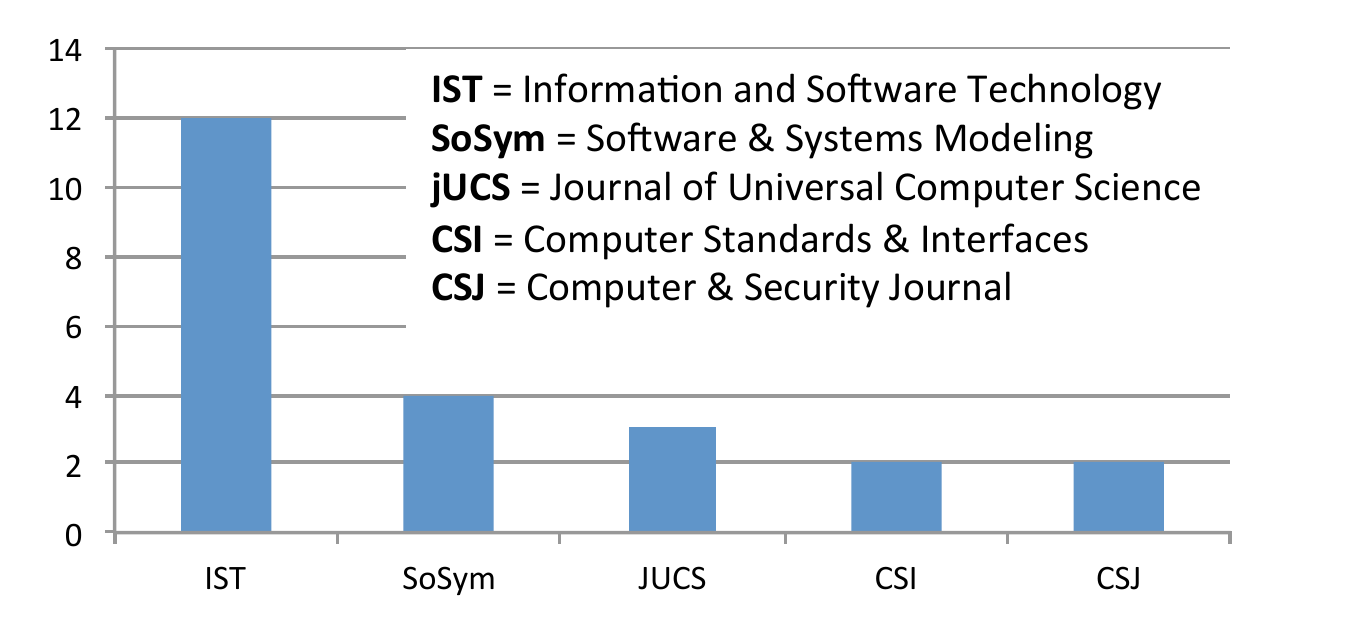}}
\hspace{3ex}
\subfloat[conferences]{\includegraphics[width=0.40\textwidth]{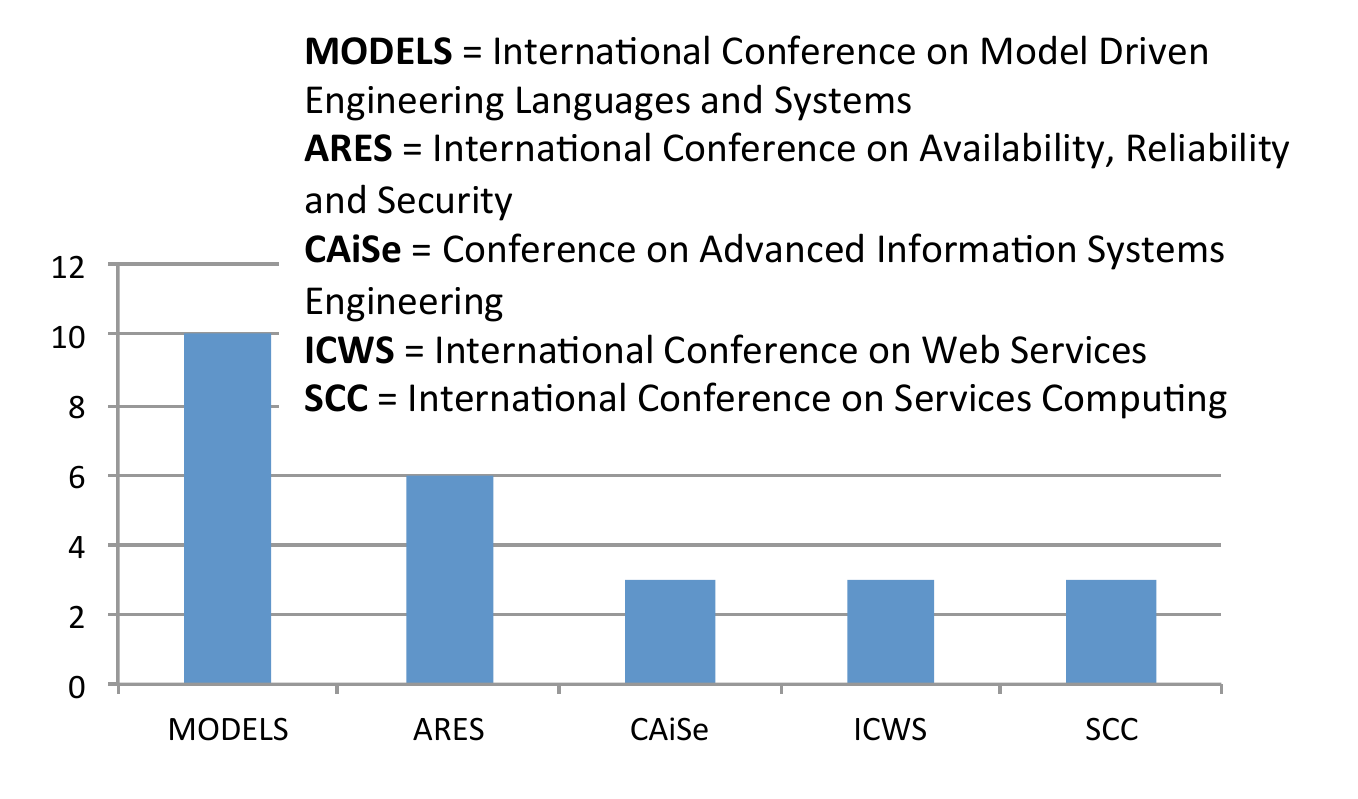}}
\hspace{3ex}
\subfloat[workshops]{\includegraphics[width=0.12\textwidth]{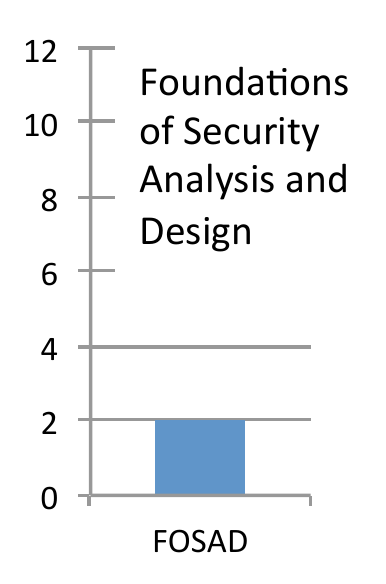}}
\caption{Number of papers for the sources with the most MDS papers found in this review}
\label{fig:mostpapers}
\end{figure*}

In terms of publication venues, Information and Software Technology (IST) journal and ACM/IEEE International Conference on Model Driven Engineering Languages and Systems (MODELS) are so far the most popular venues for publication of primary \mds papers. 
Fig. \ref{fig:mostpapers} shows that at least 10 primary \mds publications have been found in each of these two venues. 
The next two attractive venues for primary \mds papers are ARES (security conference), and SoSym (\mde journal). 
Primary \mds papers were also published at some other general journals (Journal of Universal Computer Science) or domain specific conferences (IEEE International Conference on Web Services). 
The proceedings of Tutorial Lectures on Foundations of Security Analysis and Design (FOSAD) contains some significant primary \mds approaches as well. 
In general, except ARES and CSJ, conferences and journals specific for security do not seem to be the common venues for \mds publications yet. 

%
% \begin{figure*}[htbp]
%\center
%\subfloat[journals]{\includegraphics[height=0.15\textheight]{Figs/most_journals_with2014}}
%\hspace{3ex}
%\subfloat[conferences]{\includegraphics[height=0.15\textheight]{Figs/most_conferences_with2014}}
%\hspace{3ex}
%\subfloat[workshops]{\includegraphics[height=0.15\textheight]{Figs/most_workshops_with2014}}
%\caption{Number of papers for the sources with the most MDS papers found in this review}
%\label{fig:mostpapers}
%\end{figure*}

%%%%%%%%%%%% SECTION %%%%%%%%%%%%%%%%%%%%%%
%\input{discussion.tex}

%%%%%%%%%%%% SECTION %%%%%%%%%%%%%%%%%%%%%%
%\input{validity.tex}
\section{Threats to validity}\label{sect_Validity}

We discuss the threats to validity of this \slr according to the lessons learned on validity in \slrs \cite{Kitchenham_2007} and our own experience. 

%There are several threats to validity that may affect the results of this review. 

\subsection{The search process}
To maximise the relevant articles returned by the search engines, we kept the search string not too specific but still reflecting what we wanted to search for. 
Moreover, the search string was used for searching not only in the titles, abstracts but also in the full text of an article. 
Only the search engine of Web of Knowledge (ISI) does not provide the option for searching in full text. 
This limitation could affect the search results returned by ISI. 
%Even though this step would lead to most of the papers that already found in the automatic search. 
To minimise the possibility of missing relevant papers, we kept our search string generic so that we cover as many relevant papers as possible (more than 10 thousands relevant papers found). 
To complement for the automatic search, we have also conducted the manual search on relevant journals and proceedings of relevant conferences. 
Then, to mitigate the limitations of automatic and manual searches, we have adopted the snowballing strategy. 
Even though only three out of five steps of the snowballing strategy were adopted, those are the key steps. 
Moreover, we already conducted the extensive automatic and manual searches which covered thousands of relevant publications, and resulted in a large set of primary \mds papers. 
That is why conducting only three key steps of snowballing strategy would be fair enough. 
Another possible threat is that we did not extensively search for books related to \mds. 
However, we did include the option to also search for book chapters while performing automatic search. 
In fact, we found out some book chapters that got into our final selected papers for data extraction, e.g. \cite{Architecture}, \cite{Jan2005}. 
%But we should also have searched manually for books relevant to \mds.

\subsection{Selection of primary studies}
A large part of the search and selection process was conducted by the first author. 
Some publications might have been missed. 
To mitigate this risk, every doubtful or "borderline" publication was not dismissed in the first place but rather being cross-checked and discussed by all the reviewers. 
Additionally, our clearly predefined review protocol with inclusion and exclusion criteria helped to reduce the reviewers' bias in the selection of primary studies. 

The results of this \slr papers are based on the data extracted and synthesised from the selected \mds studies. 
Note that we have applied the citation criterion to estimate the quality and impact factors of the selected primary \mds studies. 
Even though this criterion is not too strict, applying it caused a number of \mds papers not to be included. 
We realized that some of the excluded \mds papers are related to the included primary \mds studies. 
To mitigate the risk of missing some important data of the primary \mds studies, we put back the excluded \mds papers that are related to the primary \mds studies. 
In total, we re-selected 15 \mds papers as the "sidekick" papers to be included in the final set for data extraction. 

Some key selection criteria in this \slr are time-bound. 
The citation criterion for selecting primary \mds papers is based on the numbers of citations provided by Google Scholar engine. 
The selection of venues for conducting manual search is based on Microsoft Research ranking website. 
Google citations will change from time to time. 
Similarly, rankings of conferences and journals will change. 
Those time-bound metrics influence the reproduction of this \slr. 
So, some papers which were not selected as primary \mds papers because of the citation criterion would satisfy this criterion later on. 
%But this is a natural fact of any \slr. 

%\subsection{Data extraction and misclassification}

%%%%%%%%%%%% SECTION %%%%%%%%%%%%%%%%%%%%%%
%\input{relatedWork.tex}
\section{Related work}
\label{sect.relatedWork} %Related work is presented here.
%There are some related surveys (\cite{Kasal:2011:MDM:1955602.1956038}, \cite{Basin:2011:DMS:1998441.1998443}, \cite{AdvMDS}, \cite{Uzunov:jucs_18_20:engineering_security_into_distributed}), and only one \slr
%(\cite{Jensen:2011:SMD:2065363.2066253}) of \mds. 

In \cite{Kasal:2011:MDM:1955602.1956038}, the authors present a survey on \mds. 
They propose an evaluation based on the work of Khwaja and Urban \cite{doi:10.1142/S0218194002001062}. 
The study revealed that approaches that analyse implementations of modelled systems are still missing. 
Due to the fact that implementations are not generated automatically from formal specifications, verification of running
code is reasonable. 
The main drawback of \cite{Kasal:2011:MDM:1955602.1956038} is that it is not a \slr. 
As a result, there are some well-known approaches that are missing in \cite{Kasal:2011:MDM:1955602.1956038}, such as \textsc{SecureUML} \cite{Basin2006a}.

In \cite{Basin:2011:DMS:1998441.1998443}, Basin et al. went through a ``Decade of Model-Driven Security" by presenting a survey focusing on their specific \mds approach called \textsc{SecureUML}. 
The authors claim that \mds has enormous potential, mainly because Security-Design Models provide a clear, declarative, high-level language for specifying security details. 
The potential is even more, when the security models rely on a well-defined semantics. 
The main drawback of \cite{Basin:2011:DMS:1998441.1998443} is that it only considers the work around \textsc{SecureUML}.

\cite{Uzunov:jucs_18_20:engineering_security_into_distributed} is a survey of model-based security methodologies for distributed systems. 
The papers surveyed in \cite{Uzunov:jucs_18_20:engineering_security_into_distributed} are not only about model-driven methodologies but also architecture-driven methodologies, pattern-driven methodologies, and agent-driven methodologies. 
Thus the focus is not specifically \mds but rather security engineering for distributed systems in general. 
Our paper explicitly targets \mds methodologies as described in the previous sections.

In \cite{AdvMDS}, five well-known \mds approaches, i.e. UMLsec, SecureUML, Sectet, ModelSec, and SecureMDD, are summarised, evaluated, and discussed. 
These five \mds approaches are also confirmed in this paper. 
It can be seen that our \slr results are complementary to the contributions of the normal survey papers, e.g. \cite{AdvMDS}, \cite{Uzunov:jucs_18_20:engineering_security_into_distributed}. 
Those survey papers perform in depth analysis of some significant \mds approaches by elaborating one after another. 
But our \slr performs a \slr in both width and depth of \mds research which result in not only (evidently) significant \mds approaches but also emerging considerable \mds approaches. 
It is the first \mds literature review that systematically considers all relevant publications using explicit evaluation and extraction criteria.
Furthermore, our \slr provides a detailed look at all the key artefacts of any \mds approaches such as modelling techniques, security concerns, how model transformations employed, how verification and validation methods used, and case studies, and application domains. 
We also provide a trend analysis for the development of \mds research area. 

\cite{Jensen:2011:SMD:2065363.2066253} is closer to our \slr. 
%It is also a \slr on \mds. 
The authors propose three research questions with the goal to determine if the current \mds approaches focus on code generation or having empirical studies. 
The study shows that there is a need for more empirical studies on \mds (none exists), and that standardisation is key to achieve the objectives of \mdd/\mda (which are increased portability and interoperability). 
However, \cite{Jensen:2011:SMD:2065363.2066253} presents several drawbacks and differences from our paper. 
First, their search strategy is very limited compared to our three-pronged search strategy. 
Second, concerning the \slr protocol, no evaluation criteria and data extraction strategy are given. 
Moreover, their exclusion criteria are very narrow. %strict to the research questions. 
Consequently, the authors exclude significant papers in the field, e.g. \textsc{UMLsec} papers. 
Also, the authors exclude \aom approaches, because they consider that \aom does not consider security aspects as specific aspects (i.e. different from other aspects). 
Our work covers all the limitations of \cite{Jensen:2011:SMD:2065363.2066253} and provides much more extensive \slr on the topic.

%%%%%%%%%%%% SECTION %%%%%%%%%%%%%%%%%%%%%%
%\input{conclusion.tex}
\section{Conclusions}
\label{sect.conclusion}
We have presented an extensive systematic literature review on model-driven approaches for developing secure systems. 
The \slr is based on a rigorous three-pronged search process, which combined automatic search and manual search with snowballing strategy. 
Using 9 clearly predefined selection criteria, $108$ \mds papers have been strictly selected, and then reviewed. 
From these primary \mds papers, we extracted and synthesised the data to answer our research questions. 

%some security concerns are addressed much more than others
The results show that most \mds approaches focus on \emph{authorisation} and \emph{confidentiality} while only few publications address further security concerns like \emph{integrity}, \emph{availability}, and \emph{authentication}. 
% not systematically dealing with multiple security concerns.
Moreover, security concerns are often dealt with separately. 
Very few \mds approaches tackle multiple security concerns simultaneously, systematically. 
%Only 9\% of the examined \mds papers propose methodologies to tackle three key security concerns (Authentication, Authorisation, and Confidentiality) together. 
Not only multiple security concerns are less tackled, but also the inter-relations among security concerns are rarely taken into account systematically in \mds approaches. 
All these findings urge for more attention from the \mds research community to the less tackled security concerns, to the solutions that can deal with multiple security concerns simultaneously, systematically. 

% few weaving
% bit separation
Most of the approaches try to separate security concerns from core business logic, but only few weave security aspects into primary models.
% popular \uml profiles
The \uml profile mechanism is often used for the definition of security-oriented \dsl{}s, but some approaches have introduced non-\uml based \dsl{}s. 
It can be understandable that standardised, common \uml models are broadly used by \mds approaches. 
Anyway, defining \dsl{}s plays a key role in \mds because that way allows better capturing the specific semantics of security concerns. 
% but lack semantic foundation needed for formal analysis
%This may also be one of the reasons why most security modelling languages lack a thorough semantic foundation, which is needed not only for automated formal analyses. 
Still few security modelling languages are introduced with a thorough semantic foundation, which is needed for automated formal analyses. 
Most of the \mds papers use only structural models. 
Using solely one type of models could not be enough to be able to express multiple security concerns simultaneously. 

% only incomplete transformation toolchain not generating functional code and security infrastructure 
\mmt{}s and \mtt{}s are important in any \mde approach. 
In \mds, \mmt{}s and \mtt{}s are often used but implementation details and tools are not often provided. 
Many examined \mds papers do not specifically provide any implementation information about \mmt{}s. 
These papers just provide mapping rules for transforming models, or even without clearly defined transformation rules/mappings. 
Among the transformation tools provided or mentioned, not only general-purpose \mmt and \mtt tools but also many ad-hoc, specific (Java-based) tools are used. 
Developing tool chains (based on \mmts and \mtts) to derive from models to implementation code is an important piece of future work. 
Few complete tool chains to automate (most of) the \mds development process have emerged, but are very rare. 
% lack of empirical evaluation and benchmarks
Most approaches discuss illustrative examples or academic case studies but lack in-depth evaluations, e.g. using common benchmarks or empirical studies.
% summary
Altogether, our literature review shows that many \mds approaches are successful for specific, isolated security concerns or application domains, but lack formality, automation, process-integration and evaluation. 

%principal MDS approaches vs. emmerging/less common MDS approaches
In our review, we have identified 5 principal \mds studies which seem more mature than the others. 
%Each principal \mds approach is presented by at least 7 primary \mds papers in our final set of selected \mds papers. 
On the other hand, there are also other emerging/less common \mds approaches that we have found out. 
%trend analysis
Another important finding comes from the trend analysis on the key artefacts of \mds over more than a decade. 
We can see that there was a peak time of \mds publications in 2009. 
After that, the trend of primary \mds publications looks more stable. 
The IST journal and MODELS conference are so far the most popular venues for publication of \mds papers.

%Future work
%Contact authors?

%Next SLR on MDS
Our \slr protocol and the list of finally selected \mds papers could be used as the base for a follow-up \slr of \mds in the future. 
A reviewer would need to check again the citation criterion for those primary \mds papers using up-to-date citation numbers on Google Scholar. 
After obtaining a subset of \mds papers from the original set, a forward snowballing on that subset should be conducted. 
Only forward snowballing is required in this step because new, extra \mds papers will only be found in this way. 
After reviewing and selecting a new set of \mds papers from the forward snowballing step, the full snowballing process can be operated on the new set. 
The finally newly found \mds papers after snowballing will be included into the base subset to form a new final set for data extraction, synthesis, and analysis. 

%ACKNOWLEDGMENTS are optional
\section{Acknowledgments}
This work is supported by the Fonds National de la Recherche (FNR), Luxembourg, under the MITER project C10/IS/783852.

%\begin{appendix}
% \input{appendix.tex}
%\end{appendix}

% \bibliographystyle{abbrv}
% \bibliography{referencesWithList99}  % sigproc.bib is the name of the Bibliography in this case
% \setlength\bibitemsep{0.1ex} % The vertical space between the individual entries in the bibliography
% \setlength\biblabelsep{0.5ex} % The horizontal space between entries and their corresponding labels in the bibliography
\printbibliography
% ACM needs 'a single self-contained file'!
\end{document}

%% file: note1.tex
 \textit{Note}: 
 Supported (\checkmark); 
 Partially supported (o); 
 Not supported (\textsf{X}); 
 Controlled experiment (CE); 
 Industrial case study (ICS); 
 Academic case study (ACS); 
 Illustrative example (IE); 
 Not provided (NP); 

%% file: note2.tex
 Non-restrictive (NR); 
 Self-developed (Self-); 
 Endogenous (Endo); 
 Exogenous (Exo); 
 Structural (S); 
 Behavioural (B); 
 Others (O)